\documentclass{vldb}
\usepackage{graphicx}
\usepackage{epsfig}
\usepackage{algorithm}
\usepackage{algorithmic}
\usepackage{times}
\usepackage{amsfonts}
\usepackage{latexsym}
\usepackage{color}

\newtheorem{lemma}{Lemma}

\def\sharedaffiliation{%
\end{tabular}
\begin{tabular}{c}}
\begin{document}

\title{Discovery of Convoys in Trajectory Databases}
\numberofauthors{1}
\author{\hspace{-0.2cm} Hoyoung Jeung$^{\dag}$ \hspace{0.3cm} Man Lung Yiu$^{\ddag}$  \hspace{0.3cm} Xiaofang Zhou$^{\dag}$ \hspace{0.3cm} Christian S. Jensen$^{\ddag}$ \hspace{0.3cm} Heng Tao Shen$^{\dag}$
    \sharedaffiliation
    \affaddr{\hspace{0.2cm}$^{\dag}$The University of Queensland \hspace{3.3cm} $^{\ddag}$Department of Computer Science}\\
    \affaddr{\hspace{-1.0cm}National ICT Australia (NICTA), Brisbane \hspace{3.0cm} Aalborg University, Denmark} \\
    \affaddr{\hspace{-1.5cm} \{hoyoung, zxf, shenht\}@itee.uq.edu.au \hspace{3.7cm} \{mly, csj\}@cs.aau.dk} \\
}

\maketitle

\newtheorem{myproperty}{Property}
\newtheorem{mylemma}{Lemma}
\newtheorem{mytheorem}{Theorem}
\newtheorem{myproof}{Proof}
\newtheorem{mydefinition}{Definition}
\newtheorem{myexample}{Example}

\newcommand{\algorithmicinput}{\textbf{Input:}}
\newcommand{\algorithmicoutput}{\textbf{Output:}}
\newcommand{\algorithmicdescription}{\textbf{Description:}}
\newcommand{\algorithmicbreak}{\textbf{break}}
\newcommand{\algorithmiccontinue}{\textbf{continue}}
\newcommand{\algorithmicreturn}{\textbf{return~}}
\newcommand{\INPUT}{\item[{\algorithmicinput}]$\phantom{1}$\\}
\newcommand{\OUTPUT}{\item[\algorithmicoutput]$\phantom{1}$\\}
\newcommand{\DESCRIPTION}{\item[\algorithmicdescription]$\phantom{1}$\\}
\newcommand{\BREAK}{\STATE{\algorithmicbreak}}
\newcommand{\CONTINUE}{\STATE{\algorithmiccontinue}}
\newcommand{\RETURN}{\STATE{\algorithmicreturn}}

\newcommand{\goodgap}{\hspace{\subfigtopskip} \hspace{\subfigbottomskip}}

\setlength{\textfloatsep}{2ex} \setlength{\intextsep}{1.2ex}
\setlength{\dbltextfloatsep}{2ex} \addtolength{\topskip}{-2.0mm}

\begin{abstract}
As mobile devices with positioning capabilities continue to proliferate, data management for
so-called trajectory databases that capture the historical movements of populations of moving
objects becomes important. This paper considers the querying of such databases for convoys, a
convoy being a group of objects that have traveled together for some time.

More specifically, this paper formalizes the concept of a convoy query using density-based notions,
in order to capture groups of arbitrary extents and shapes. Convoy discovery is relevant for
real-life applications in throughput planning of trucks and carpooling of vehicles. Although there
has been extensive research on trajectories in the literature, none of this can be applied to
retrieve correctly exact convoy result sets. Motivated by this, we develop three efficient
algorithms for convoy discovery that adopt the well-known filter-refinement framework. In the
filter step, we apply line-simplification techniques on the trajectories and establish distance
bounds between the simplified trajectories.
%
This permits efficient convoy discovery over the simplified trajectories without missing any actual
convoys. In the refinement step, the candidate convoys are further processed to obtain the actual
convoys.
%
Our comprehensive empirical study offers insight into the properties of the paper's proposals and
demonstrates that the proposals are effective and efficient on real-world trajectory data.
\end{abstract}

\section{Introduction} \label{sec:intro}
Although the mobile Internet is still in its infancy, very large
volumes of position data from moving objects are already being
accumulated. For example, Inrix, Inc. based in Kirkland, WA receive
real-time GPS probe data from more than 650,000 commercial fleet,
delivery vehicles, and taxis~\cite{inrix}.
As the mobile Internet continues to proliferate and as congestion becomes increasingly widespread
across the globe, the volumes of position data being accumulated are likely to soar. Such data may
be used for many purposes, including travel-time prediction, re-routing, and the identification of
ride-sharing opportunities.
This paper addresses one particular challenge to do with the
extraction of meaningful and useful information from such position
data in an efficient manner.

\begin{figure}[hbt]
    \centering
    \includegraphics[width=6.0cm]{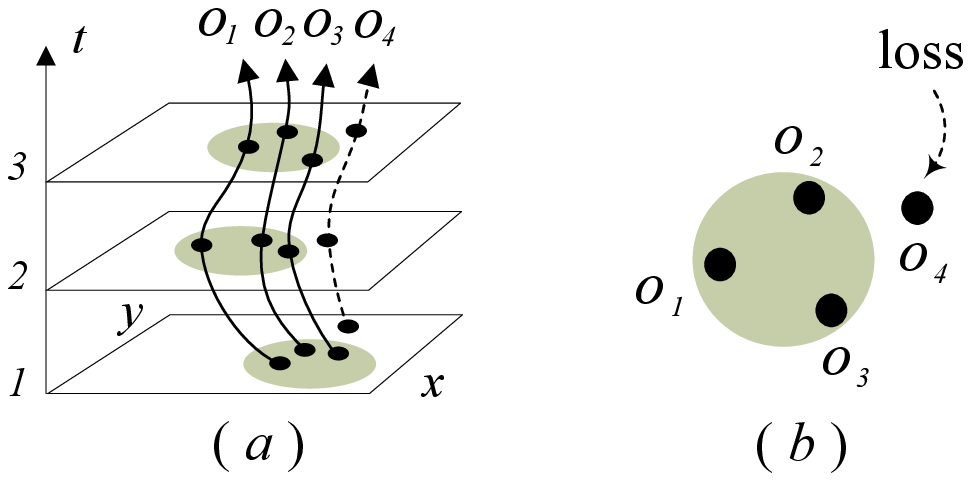}
    \caption{\textit{Lossy-flock} Problem}
    \label{fig:lossyflock}
\end{figure}

The movement of an object is given by a continuous curve in the $(\mbox{space}, \mbox{time})$
domain, termed a {\em trajectory}. The past trajectory of an object is typically approximated based
on a collection of time-stamped positions, e.g., obtained from a GPS device. As an example,
Figure~\ref{fig:lossyflock}(\emph{a}) depicts the trajectories of four objects $o_1$, $o_2$, $o_3$,
and $o_4$ in $(x,y,t)$ space.

Given a collection of trajectories, it is of interest to discover
groups of objects that travel together for more than some minimum
duration of time.
A number of applications may be envisioned. The identification of delivery trucks with coherent
trajectory patterns may be used for throughput planning. The discovery of common routes among
commuters may be used for the scheduling of collective transport. The identification of cars that
follow the same routes at the same time may be used for the organization of carpooling, which may
reduce congestion, pollution, and CO$_2$ emissions.

The discovery of so-called \emph{flocks}
\cite{st-flock4,st-flock3,st-flock2} has received some attention. A
flock is a group of objects that move together within a disc of some
user-specified size.
On the one hand, the chosen disk size has a substantial effect on the
results of the discovery process. On the other hand, the selection of
a proper disc size turns out to be difficult, as situations can occur
where objects that intuitively belong together or do not belong
together are not quite within any disk of the given size or are within
such a disk. And for some data sets, no single appropriate disc size
may exist that works well for all parts of the $(\mbox{space},
\mbox{time})$ domain.
In Figure~\ref{fig:lossyflock}(\emph{a}), all objects travel together in a natural group.  However,
as shown in Figure~\ref{fig:lossyflock}(\emph{b}), object $o_4$ does not enter the disc and is not
discovered as a member of the flock. A key reason why this \emph{lossy-flock} problem occurs is
that what constitutes a flock is very sensitive to the user-specified disc size, which is
independent of the data distribution.
In addition, the use of a circular shape may not always be appropriate. For example, suppose that
two different groups of cars move across a river and each group has a long linear form along roads.
A sufficient disc size for capturing one group may also capture the other group as one flock.
Ideally, no particular shape should be fixed apriori.

To avoid rigid restrictions on the sizes and shapes of the trajectory patterns to be discovered, we
propose the concept of \emph{convoy} that is able to capture generic trajectory pattern of any
shape and any extent. This concept employs the notion of density connection \cite{s-clustering3},
which enables the formulation of arbitrary shapes of groups.
Given a set of trajectories $O$, an integer $m$, a distance value $e$, and a lifetime $k$, a
\emph{convoy query} retrieves all groups of objects, i.e., \emph{convoys}, each of which has at
least $m$ objects so that these objects are so-called density--connected with respect to distance
$e$ during $k$ consecutive time points. Intuitively, two objects in a group are density--connected
if a sequence of objects exists that connects the two objects and the distance between consecutive
objects does not exceed $e$. (The formal definition is given in Section~\ref{sec:prob}.)
Each group of objects in the result of a convoy query is associated
with the time intervals during which the objects in the group traveled
together.

The efficient discovery of convoys in a large trajectory database is
a challenging problem. Convoy queries compute sets of objects and
are more expensive to process than spatio-temporal
joins~\cite{st-join4}, which compute pairs of objects.
Past studies on the retrieval of similar trajectories generally use
distance functions that consider the distances between pairs of
trajectories across all of time~\cite{EDR,LCSS,DTW}. In contrast, we
consider distances during relatively short durations of time.
Other relevant work concerns the clustering of moving objects
\cite{st-clustering4,st-clustering1,st-clustering2}. In these works, a moving cluster exists if a
shared set of objects exists across adjacent time, but objects may join and leave a cluster during
the cluster's lifetime. Hence, moving clusters carry different semantics and do not necessarily
qualify as convoys.
%

Jeung et. al. first proposed the convoy query and outlined preliminary techniques for convoy
discovery \cite{convoy08}. In this paper, we extend the work, which develops more advanced
algorithms and analyzes each discovery method in real world settings. Specifically, we introduce
four effective and efficient algorithms for answering the convoy query. The first method adopts the
solution for moving cluster discovery to our convoy problem. The second method, called CuTS
(\emph{\underline{C}onvoy Discovery \underline{u}sing \underline{T}rajectory
\underline{S}implification}), employs the filter-refinement framework --- a set of candidate
convoys are retrieved in the filter step, and then they are further processed in the refinement
step to produce the actual convoys. In the filter step, we apply line simplification techniques
\cite{DP} on the trajectories to reduce their sizes; hence, it becomes very efficient to search for
convoys over simplified trajectories. We establish distance bounds between simplified trajectories,
in order to ensure that no actual convoy is missing from the candidate convoy set. The third method
(CuTS+) accelerates the process of trajectory simplification of CuTS to increase the efficiency of
the filter step even further. The last method, named CuTS*, is an advanced version of CuTS that
enhances the effectiveness of the filter step by introducing tighter distance bounds for simplified
trajectories.

 The main novelties of this paper are summarized as follows:

\begin{itemize}
\setlength{\parskip}{-3pt} \item
Our filter step operates on trajectories processed by line simplification techniques; this is
different from most related works that employ spatial approximation (e.g., bounding boxes) in the
filter step. The rationale is that conventional methods using bounding boxes introduce substantial
empty space, rendering them undesirable for the processing of trajectory data.

\setlength{\parskip}{-3pt} \item
To guarantee correct convoy discovery, we establish distance bounds for range search over
simplified trajectories. In contrast, the distance bounds studied elsewhere \cite{st-simple2} are
applicable only to specific query types, not to the convoy problem.

\setlength{\parskip}{-3pt} \item
We study various trajectory simplification techniques in conjunction with different query
processing mechanisms. In addition, we show how to tighten the distance bounds.

\setlength{\parskip}{-3pt} \item
We present comprehensive experimental results using several real trajectory data sets, and we
explain the advantages and disadvantages of each proposed method.

\end{itemize}

The remainder of this paper is organized as follows: In Section~\ref{sec:related}, we discuss
previous methods related to the convoy query. We formulate the focal problem of this paper in
Section~\ref{sec:prob}. A modified method of moving cluster for the convoy discovery is shown in
Section~\ref{sec:cmc}. We propose more efficient methods based on trajectory simplification in
Sections~\ref{sec:cuts} and \ref{sec:ext}. Section~\ref{sec:exp} reports the results of
experimental performance comparisons, followed by conclusions in Section~\ref{sec:conc}.

\section{Related Work} \label{sec:related}

We first review existing work on trajectory clustering and, then cover
trajectory simplification, which is an important aspect of our
techniques for convoy discovery. We end by considering spatio-temporal
joins and distance measures for trajectories.

\subsection{Clustering over Trajectories} \label{sec:trjclust}

\begin{figure}[hbt]
    \centering
    \includegraphics[width=6.5cm]{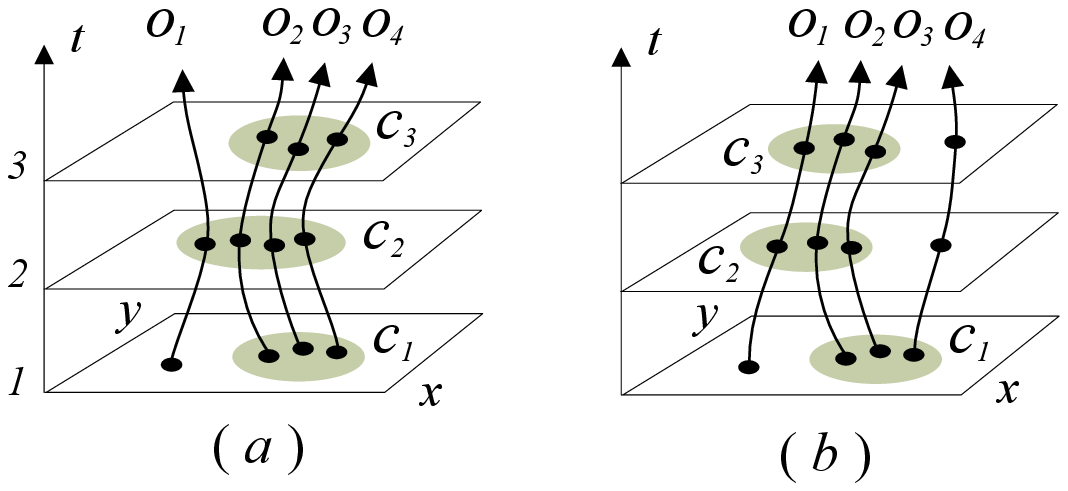}
    \caption{Convoys Versus Moving Clusters}
    \label{fig_lossymc}
\end{figure}

Given a set of points, the goal of {\em spatial clustering} is to
form clusters (i.e., groups) such that (i) points within
the same cluster are close to each other, and (ii) points from
different clusters are far apart.
In the context of trajectories, the locations of trajectories can be clustered at chosen time
points. Consider the trajectories in Figure~\ref{fig_lossymc}(\emph{a}). We first obtain a cluster
$c_1$ at time $t=1$, then a cluster $c_2$ at $t=2$, and eventually a cluster $c_3$ at $t=3$.

Kalnis et al. propose the notion of a {\em moving cluster} \cite{st-clustering1}, which is a
sequence of spatial clusters appearing during consecutive time points, such that the portion of
common objects in any two consecutive clusters is not below a given threshold parameter $\theta$,
i.e., $\frac{ |c_t \cap c_{t+1}| }{ |c_t \cup
  c_{t+1}| } \ge \theta$, where $c_t$ denotes a cluster at time $t$.
There is a significant difference between a convoy and a moving
cluster.
For instance, in Figure~\ref{fig_lossymc}(\emph{a}), $o_2,o_3$, and $o_4$ form a convoy with 3
objects during 3 consecutive time points. On the other hand, if we set $\theta=1$ (i.e., require
100\% overlapping clusters), the overlap between $c_1$ and $c_2$ is only $\frac{3}{4}$, and the
above objects will not be discovered as a moving cluster. Next, in
Figure~\ref{fig_lossymc}(\emph{b}), if we set $\theta = \frac{1}{2}$ then $c_1$, $c_2$, and $c_3$
become a moving cluster. However, this is not a convoy.

Spiliopoulou et al.~\cite{st-clustering9} study transitions in moving clusters (e.g., disappearance
and splitting) between consecutive time points. As transitions are based on the consideration of
common objects at consecutive time points, their techniques do not support convoy discovery either.
Next, Li et al.~\cite{st-clustering2} study the notion of \emph{moving micro cluster}, which is a
group of objects that are not only close to one another at the current time, but are also expected
to move together in the near future.
Recently, Lee et al.~\cite{TRACLUS} have proposed to partition trajectories into line segments and
build groups of close segments. This proposal does not consider the temporal aspects of the
trajectories. As a result, some objects can belong to the same group even though they have never
traveled close together (at the same time). Most recently, Jensen et al.~\cite{st-clustering4} have
proposed techniques for maintaining clusters of moving objects. They consider the clustering of the
current and near-future positions, while we consider past trajectories.

As mentioned earlier, several slightly different notions of a \emph{flock}
\cite{st-flock3,st-flock2} relate to that of a convoy.  The notion most relevant to our study
defines a flock as a group of at least $m$ objects staying together within a circular region of
radius $e$ during a specified time interval \cite{st-flock4,st-flock3}. Al-Naymat et
al.~\cite{st-flock4} apply random projection to reduce the dimensionality of the data and thus
obtain better performance. Gudmundsson et al.~\cite{st-flock3} propose approximation techniques and
exploit an index to accelerate the computation of flock. It is also shown that the discovery of the
longest-duration flock is an NP-hard problem. It is worth noticing that these studies exhibit the
lossy-flock problem identified in Section~\ref{sec:intro}.

\subsection{Trajectory Simplification} \label{sec:trjsimple}
A trajectory is often represented as a {\em polyline}, which is a sequence of connected line
segments. Line simplification techniques have been proposed to simplify polylines according to some
user-specified resolution \cite{DP,DP-speed}.

The Douglas-Peucker algorithm (DP) \cite{DP} is a well-known and efficient method among the line
simplification techniques. Given a polyline specified by a sequence of $T$ points $\langle p_1,
p_2, \cdots, p_T \rangle$ and a distance threshold $\delta$, the goal is to derive a new polyline
with fewer points while deviating from the original polyline by at most $\delta$. The DP algorithm
initially constructs the line segment $\overline {p_1 p_T}$. It then identifies the point $p_i$
farthest from the line. If this point's (perpendicular) distance to the line is within $\delta$
then DP returns $\overline {p_1 p_T}$ and terminates. Otherwise, the line is decomposed at $p_i$,
and DP is applied recursively to the sub-polylines $\langle p_1, p_2, \cdots, p_i \rangle$ and
$\langle p_i,\cdots, p_T \rangle$.
As the worst-case time complexity of this algorithm is $O(T^2)$,
%
Hershberger et al.~\cite{DP-speed} show a faster version of this method with time complexity of
$O(T \cdot \log T)$. However, it is assumed that an object's trajectory cannot intersect itself,
which is not a valid assumption for the data we consider.

\begin{figure}[hbt]
    \centering
    \includegraphics[width=8.5cm]{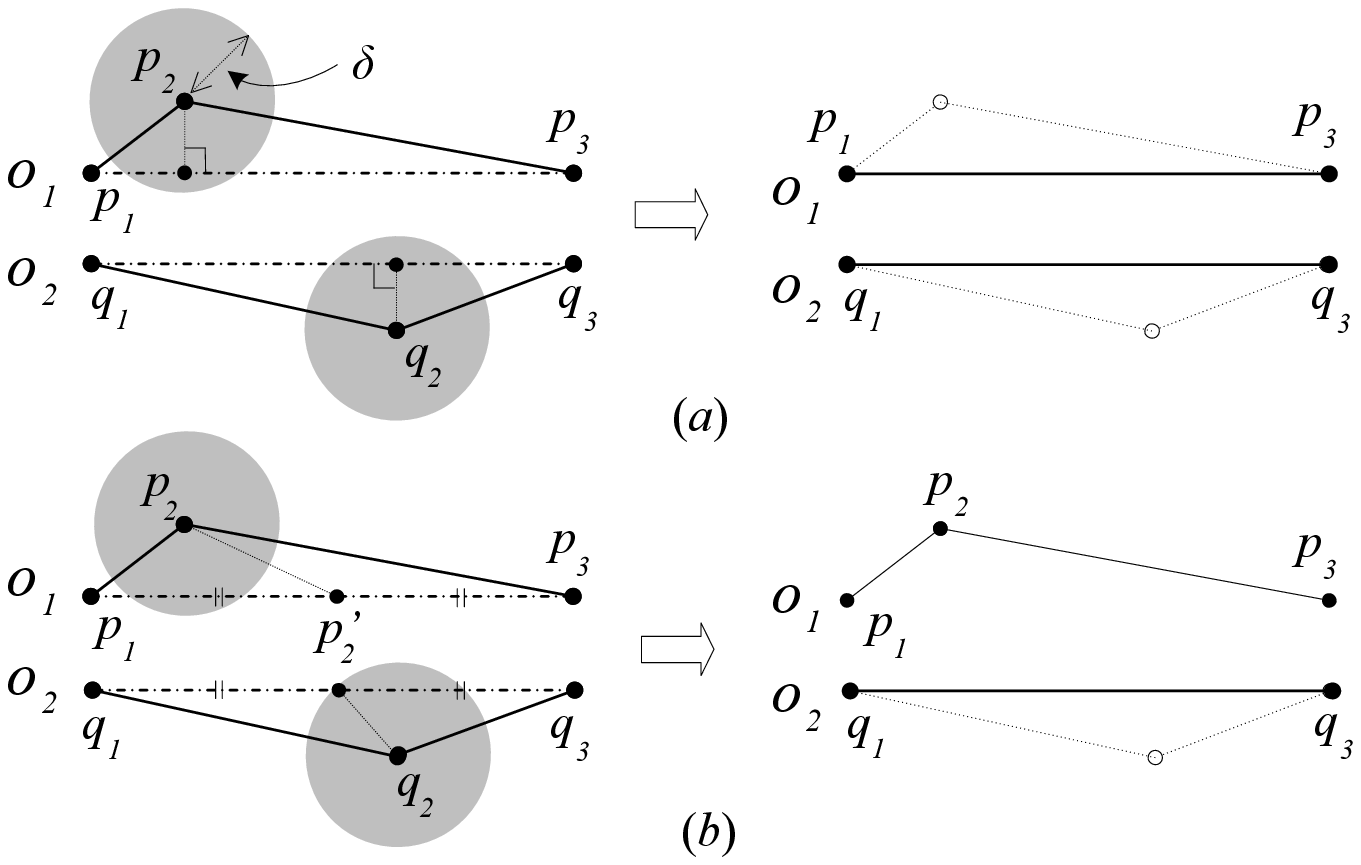}
    \caption{Comparison of Different Trajectory Simplifications}
    \label{fig_stsimpleFT}
\end{figure}

The DP technique deals with line simplification only in the spatial domain, ignoring the time
domain of the trajectories. Consider the example in Figure~\ref{fig_stsimpleFT}(\emph{a}).
Since the distance from $p_2$ to $\overline {p_1p_3}$ is within $\delta$, the DP algorithm omits
$p_2$ and simply returns $\overline {p_1p_3}$. Similarly, $q_2$ is also omitted and the polygon is
simplified to $\overline {q_1q_3}$.

In contrast, Meratnia et al.~\cite{st-simple1} take into account the temporal aspects in line
simplification. Figure~\ref{fig_stsimpleFT}(\emph{b}) exemplifies the working procedure of their
algorithm (say, DP*). First, DP* derives the point $p'_2$ on the line $\overline {p_1p_3}$ by
calculating the ratio of $p_2$'s time between $t$=1 of $p_1$ and $t$=3 of $p_3$. Then, it measures
the distance $D(p_2,p'_2)$ between $p_2$ and $p'_2$, instead of the perpendicular distance from
$p_2$ to $\overline {p_1p_3}$. Since $D(p_2,p'_2)> \delta$, $p_2$ is still kept after the
simplification, while it was removed by using DP in Figure~\ref{fig_stsimpleFT}(\emph{a}).
%
%

\subsection{Distance Measures and Joins} \label{sec:stjoin}

A basic way of measuring the distance between two trajectories used in the literature is to compute
the sum of their Euclidean distances over time points. Such a distance measure may not be able to
capture the inherent distance between trajectories because it does not take into account particular
features of trajectories (e.g., noise, time distortion). Thus, it is important to devise a distance
function that ``understands'' the characteristics of trajectories.

A well-known approach is Dynamic Time Warping (DTW) \cite{DTW}, which applies dynamic programming
for aligning two trajectories in such a way that their overall distance is minimized.
More recent proposals for trajectory distance functions include
Longest Common Subsequence (LCSS)~\cite{LCSS}, Edit Distance on Real
Sequence (EDR)~\cite{EDR}, and Edit distance with Real Penalty
(ERP)~\cite{ERP}.
Lee et al.~\cite{TRACLUS} point out that the above distance
measures capture the global similarity between two trajectories, but
not their local similarity during a short time interval. Thus, these
measures cannot be applied in a simple manner for convoy discovery.

Given two data sets $P_1$ and $P_2$, spatio-temporal joins find pairs
of elements from the two sets that satisfy predicates with both
spatial and temporal attributes~\cite{st-join2}. The {\em close-pair
  join} reports all object pairs ($o_1$, $o_2$) from $P_1 \times P_2$
with distance $D_{\tau}(o_1, o_2)\leq e$ within a time interval $\tau$
being bounded by a user-specified distance $e$. Plane-sweep techniques
\cite{st-join6,st-join5} have been proposed for evaluating
spatio-temporal joins. Like the close-pair join, the {\em trajectory
  join} \cite{st-join4} aims at retrieving all pairs of
similar trajectories between two datasets. Bakalov et al.~\cite{st-join4} represent trajectories as
sequences of symbols and apply sliding window techniques to measure the symbolic distance between
possible pairs. These studies consider pairs of objects, whereas we consider sets of objects.

\section{Problem Definition}  \label{sec:prob}

This section formalizes the convoy problem. We start with the definitions of distances for points,
line segments, and bounding boxes :
\begin{mydefinition}
  \textbf{(Distance Functions)}\\
  \vspace{-0.4cm}
\begin{itemize}
\setlength{\parskip}{-2pt} \item Given two points $p_u$ and $p_v$,
  $D(p_u,p_v)$ is defined as the Euclidean distance between $p_u$ and
  $p_v$.
\setlength{\parskip}{-2pt} \item Given a point $p$ and a line segment $l$, $D_{PL}(p,l)$ is defined
  as the shortest (Euclidean) distance between $p$ and any point on
  $l$.
\setlength{\parskip}{-2pt} \item Given two line segments $l_u$ and $l_v$, $D_{LL}(l_u,l_v)$ is
  defined as the shortest (Euclidean) distance between any two points
  on $l_u$ and $l_v$, respectively.
\setlength{\parskip}{-2pt} \item   With $\mathcal{B}_u$ and $\mathcal{B}_v$ being boxes then
  $D_{min}(\mathcal{B}_u,\mathcal{B}_v)$ is defined as the minimum
  distance between any pair of points belonging to each of the two
  boxes.
\end{itemize}
\end{mydefinition}

The boxes introduced in the definition will be used for the bounding of line segments.
Next, the time domain is defined as the ordered set $\{t_1, t_2, \cdots, t_T\}$, where $t_j$ is a
time point and $T$ is the total number of time points.

In our problem setting, we consider a practical trajectory database model. We assume each
trajectory may have a different length from others and may also appear or disappear at any time in
$T$. In addition, each location of a trajectory can be sampled either regularly (e.g., every
second) or irregularly (i.e, some missing time points from $T$ may exist between two consecutive
time points of the trajectory).


The trajectory of an object $o$ is represented by a polyline that is given as a sequence of
timestamped locations $o=\langle p_a,p_{a+1}, \cdots, p_b \rangle$, where $p_j = (x_j,y_j,t_j)$
indicates the location of $o$ at time $t_j$, with $t_a$ being the start time and $t_b$ being the
end time. The time interval of $o$ is $o.\tau = [t_a,t_b]$. A shorthand notation is to use $o(t_j)$
for referring to the location of $o$ at time $t_j$ (i.e., location $p_j$).

\begin{figure}[hbt]
    \centering
    \includegraphics[width=5.7cm]{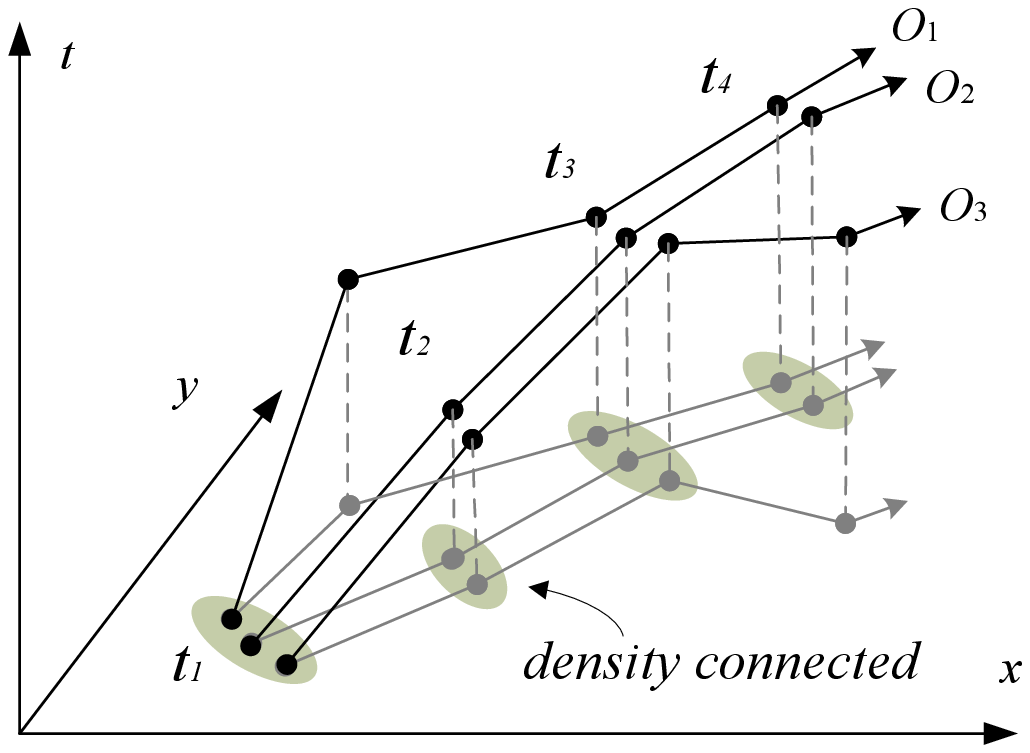}
    \caption{An Example of a Convoy}
    \label{fig_convoy}
\end{figure}

Figure~\ref{fig_convoy} illustrates the polylines representing the
trajectories of three objects $o_1, o_2$, and $o_3$, during the time
interval from $t_1$ to $t_4$.

As a precursor to defining the convoy query, we need to understand
the notion of density connection~\cite{s-clustering3}. Given a
distance threshold $e$ and a set of points $S$, the $e$-neighborhood
of a point $p$ is given as $\mathit{NH}_e(p) = \{q \in S \; | \;
D(p, q) \leq e$\}.
Then, given a distance threshold $e$ and an integer $m$, a point $p$ is {\em directly
density--reachable} from a point $q$ if $p \in \mathit{NH}_e(q)$ and $|\mathit{NH}_e(q)| \geq m$.
A point $p$ is said to be {\em density--reachable} from a point $q$ with respect to $e$ and $m$ if
there exists a chain of points $p_1$, $p_2$, ..., $p_n$ in set $S$ such that $p_1 = q$, $p_n = p$,
and $p_{i+1}$ is directly density--reachable from $p_i$.

\begin{mydefinition} \label{def:dcon} \textbf{(Density--Connected)}
  Given a set of points $S$, a point $p \in S$ is density--connected to
  a point $q \in S$ with respect to $e$ and $m$ if there exists a
  point $x \in S$ such that both $p$ and $q$ are density--reachable
  from $x$.
\end{mydefinition}

%

The definition of density--connection permits us to capture a group of ``connected'' points with
arbitrary shape and extent, and thus to overcome the the lossy-flock problem shown in Figure
\ref{fig:lossyflock}. By considering density--connected objects for consecutive time points, we
define the {\em convoy query} as follows:

\begin{mydefinition}
  \textbf{(Convoy Query)} Given a set of trajectories of $N$ objects,
  a distance threshold $e$, an integer $m$, and an integer lifetime
  $k$, the convoy query returns all possible groups of objects, so
  that each group consists of a (maximal) set of density-connected
  objects with respect to $e$ and $m$ during at least $k$ consecutive
  time points.
\end{mydefinition}


Consider the convoy query with the parameters $m=2$ and $k=3$ issued over the trajectories in
Figure~\ref{fig_convoy}. $\langle{o_2,o_3},[t_1, t_3]\rangle$ is the result, meaning that $o_2$ and
$o_3$ belong to the same convoy during consecutive time points from $t_1$ to $t_3$.

Table~\ref{tab:summary-notation} offers the notations introduced in this section and to be used
throughout the paper.
%

\begin{table} [hbt]
  \centering
  \small
  \begin{tabular}{|c|c|}
  \hline
  \textbf{Symbol}  & \textbf{Meaning} \\ \hline
   $p$        & Point/location (in the spatial domain) \\ \hline
   $t$        & Time point       \\ \hline
   $o_i$        & Original trajectory of an object \\ \hline
   $o_i(t)$   & Location of $o_i$ at time $t$ \\ \hline
   $o'_i$        & Simplified trajectory (of $o_i$) \\ \hline
   $l'_i$     & Line segment of $o'_i$ \\ \hline
   $o'_i.\tau$     & Time interval of $o'_i$ \\ \hline
   $l'_i.\tau$     & Time interval of $l'_i$ \\ \hline
   $D(p_u,p_v)$     &  Euclidean distance between points \\ \hline
   $D_{PL}(p,l)$     & The shortest distance from point to line segment \\ \hline
   $D_{LL}(l_u,l_v)$   &  The shortest distance between line segments \\ \hline
   $\mathcal{B}(l)$   &  The minimum bounding box of $l$ \\ \hline
   $D_{min}(\mathcal{B}_u,\mathcal{B}_v)$   &  The minimum distance between two boxes \\ \hline
  \end{tabular}
\caption{Summary of Notation}\label{tab:summary-notation}
\end{table}

\section{Coherent Moving Cluster (CMC)} \label{sec:cmc}

\begin{figure}[hbt]
    \centering
    \includegraphics[width=8.2cm]{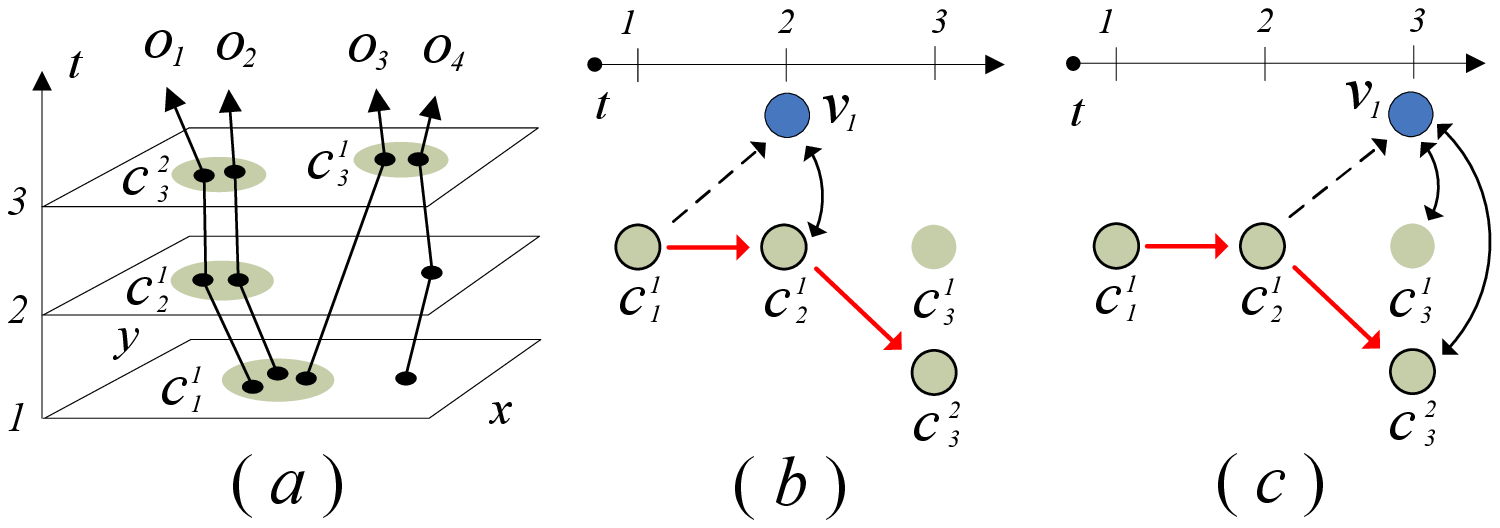}
    \caption{Query Processing of CMC, $m=2$}
    \label{fig_cmc}
\end{figure}

A simple technique for computing a convoy is to first perform (density--connected) clustering on
the objects at each time and then to extract their common objects in an attempt to form convoys.
This approach is similar to the methods for discovering moving clusters \cite{st-clustering1}.
However, those are unable to discover the exact convoy results, as explained next:

\pagebreak

\begin{itemize}

\setlength{\parskip}{-2pt} \item Let $c_t$ and $c_{t+1}$ be (snapshot) clusters at times $t$ and
  $t+1$. These clusters belong to the same \emph{moving cluster} if
  they share at least the fraction $\theta$ objects ($|c_t \cap
  c_{t+1}|/|c_t \cup c_{t+1}| \geq \theta$), where $\theta$ is a
  user-specified threshold value between 0 and 1.
 The problem of applying moving cluster methods for convoy discovery is that no absolute $\theta$ value exists that can be used to compute the exact
  convoy results---either false hits may be found, or actual convoys
  may remain undiscovered, as explained in Section~\ref{sec:trjclust}.
%

\setlength{\parskip}{-2pt} \item A moving cluster can be formed as long as two snapshot clusters
  have at least $\theta$ overlap, even for only two consecutive
  time. The lifetime ($k$) constraint does not apply to moving
  clusters, but is essential for a convoy.

\setlength{\parskip}{-2pt} \item As pointed in the previous section, a trajectory may have some
missing time points due to irregular location sampling (e.g., $o_3$ at $t=2$ in Figure
~\ref{fig_cmc}(\emph{a})). In this case, we cannot measure the density--connection for all objects
involved over those missing times.
\end{itemize}

In order to solve the above problems for convoy discovery, we extend the moving cluster method into
our \emph{Coherent Moving Cluster} algorithm (CMC). First, we generate virtual locations for the
missing time points. If any trajectory has a location at time $t_i$, but another does not during
its time interval, we apply linear interpolation to create the virtual points at $t_i$. Second, to
accommodate the lifetime ($k$) constraint, we require each candidate convoy to have (at least) $k$
clusters $c_t, c_{t+1}, \cdots, c_{t+k-1}$ during consecutive time points. Third, we test the
condition $|c_t \cap c_{t+1} \cap \cdots \cap c_{t+k-1}| \geq m$, to determine whether sufficiently
many common objects are shared. If all conditions are satisfied, the candidate is reported as an
actual convoy.

We proceed to illustrate algorithm CMC using Figure~\ref{fig_cmc},
with the parameters $m=2$ and $k=3$. Let $c_t^i$ be the $i$-th
snapshot cluster at time $t$. Clusters at time $t$ are obtained by
applying a snapshot density clustering algorithm (e.g., DBSCAN
\cite{s-clustering3}) on the objects' locations at time $t$.

Table~\ref{tab:steps-cmc} illustrates the execution steps of the algorithm. At time $t_1$, we
obtain a cluster $c_1^1$ (with objects $o_1$, $o_2$, and $o_3$) and consider it a convoy candidate
$v_1$. At time $t_2$, we retrieve a cluster $c_2^1$, which is then compared with $v_1$. Since
$c_2^1$ and $v_1$ have $m=2$ common objects, we compute their intersection and update candidate
$v_1$. At time $t_3$, we discover two clusters $c_3^1$ and $c_3^2$.  Since $c_3^1$ shares no
objects with $v_1$, we consider $c_3^1$ as another convoy candidate $v_2$. As $c_3^2$ shares $m=2$
common objects with $v_1$, we update $v_1$ to be its intersection with $c_3^2$. Eventually, $v_1$
is reported as a convoy because it contains $m=2$ common objects from clusters during $k=3$
consecutive time points.

\begin{table} [hbt]
  \centering
  \begin{tabular}{|c|c|c|}
  \hline
  \textbf{Timestamp}  & \textbf{Clusters} & \textbf{Candidate set $V$} \\ \hline
   $t_1$        & $c_1^1$           & $v_1=c_1^1$   \\ \hline
   $t_2$        & $c_2^1$           & $v_1=c_1^1 \cap c_2^1$  \\ \hline
   $t_3$        & $c_3^1$, $c_3^2$  & $v_1=c_1^1 \cap c_2^1 \cap c_3^2$, $v_2=c_3^1$  \\ \hline
  \end{tabular}
\caption{Execution Steps of CMC}\label{tab:steps-cmc}
\end{table}

Algorithm~\ref{alg:cmc} presents the pseudocode for the CMC algorithm. The algorithm takes as
inputs a set of object trajectories $O$ and convoy query parameters $m$, $k$, and $e$.

We use $V$ to represent the set of convoy candidates.  We then perform processing for each time
point (in ascending order).
The set $V_{next}$ introduced in Line~3 is used to store candidates produced at the current time
$t$. Then, we consider only objects $o \in O$ whose time intervals cover time $t$, i.e., $t \in
o.\tau$. Their locations $o(t)$ are inserted into the set $O_t$. If any object $o \in O_t$ has a
missing location at $t$, a virtual point is computed and then inserted.

Next, we apply DBSCAN on $O_t$ to obtain a set $C$ of clusters (Line~7).
%
%
The clusters in $C$ are compared to existing candidates in $V$. If they share at least $m$ common
objects (Line~11), the current objects of the candidate $v$ are replaced by the common objects
between $c$ and $v$ and are then inserted into the set $V_{next}$ (Lines~13--15). At the same time,
we increment the lifetime of the candidate (Lines~14). Each candidate with its lifetime (at least)
$k$ is reported as a convoy (Lines~17--18).

Clusters (in $C$) having insufficient intersections with existing candidates are inserted as new
candidates into $V_{next}$ (Lines~19--23). Then all candidates in $V_{next}$ are copied to $V$ so
that they are used for further processing in the next iteration.


\begin{algorithm}[!h]
\small \caption{\bf CMC (Set of object trajectories $O$, Integer $m$, Integer $k$, Distance
threshold $e$)}
\begin{algorithmic}[1]
    \STATE $V \leftarrow \emptyset$ 
    \FOR {each time $t$ (in ascending order)}
        \STATE $V_{next} \leftarrow \emptyset$
        \STATE $O_t \leftarrow \{ o(t) \; | \; o \in O \; \wedge \; t \in o.\tau   \}$
        \IF { $O_t$.size $< m $}
            \STATE skip this iteration
        \ENDIF
        \STATE $C \leftarrow$ DBSCAN($O_t, e, m$)
        \FOR {each convoy candidate $v \in V$}
            \STATE $v.$assigned $\leftarrow$ false
            \FOR {each snapshot cluster $c \in C$}
                \IF { $|c \cap v| \geq m $}
                    \STATE $v$.assigned $\leftarrow$ true
                    \STATE $v \leftarrow c \cap v$
                    \STATE $v$.endTime $ \leftarrow t$
                    \STATE $V_{next} \leftarrow V_{next} \cup v$
                    \STATE $c$.assigned $\leftarrow$ true
                \ENDIF
            \ENDFOR
            \IF {$v$.assigned = false and $v$.lifetime$\ge k$}
                \STATE  $V_{result} \leftarrow V_{result} \cup v$
            \ENDIF
        \ENDFOR
        \FOR {each $c \in C$}
            \IF {$c$.assigned = false} 
                \STATE $c$.startTime $\leftarrow t$
                \STATE $c$.endTime $\leftarrow t$
                \STATE $V_{next} \leftarrow V_{next} \cup c$
            \ENDIF
        \ENDFOR
        \STATE $V \leftarrow V_{next}$;
     \ENDFOR
     \RETURN $V_{result}$
\end{algorithmic}
\label{alg:cmc}
\end{algorithm}

\section{Convoy Discovery Using Trajectory Simplification (CuTS)} \label{sec:cuts}

The CMC algorithm incurs high computational cost because it generates virtual locations for all
missing time points and performs expensive clustering at every time. In this section, we apply the
filter-and-refinement paradigm with the purpose of reducing the overall computational cost. For the
filter step, we simplify the original trajectories and apply clustering on the simplified
trajectories to obtain convoy candidates. The goal is to retrieve a superset of the actual convoys
efficiently. In the refinement step, we consider each candidate convoy in turn. In particular, we
perform clustering on the original trajectories of the objects involved to determine whether the
convoy indeed qualifies. The resulting CuTS algorithm is guaranteed to return correct convoy
results.

\subsection{Simplifying Trajectories} \label{sec:simple}
Given a trajectory represented as a polyline $o= \langle p_1, p_2,\cdots, p_T \rangle$, and a
tolerance $\delta$, the goal of {\em trajectory simplification} is to derive another polyline $o'$
such that $o'$ has fewer points and deviates from $o$ by at most $\delta$. We say that $o'$ is a
{\em simplified trajectory} of $o$ with respect to $\delta$.

We apply the Douglas-Peucker algorithm (DP), as discussed in Section~\ref{sec:trjsimple}, to
simplify a trajectory.
Initially, DP composes the line $\overline {p_1p_T}$ and finds the point $p_i \in o$ farthest from
the line. If the distance $D_{PL}(p_i,\overline{p_1p_T}) \leq \delta$, segment $\overline {p_1p_T}$
is reported as the simplified trajectory $o'$.  Otherwise, DP recursively processes the
sub-trajectories $\langle p_1, \cdots, p_i \rangle$ and $\langle p_i,\cdots, p_T \rangle$,
reporting the con\-ca\-tenation of their simplified trajectories as the simplified trajectory~$o'$.

\begin{figure}[hbt]
    \centering
    \includegraphics[width=7.5cm]{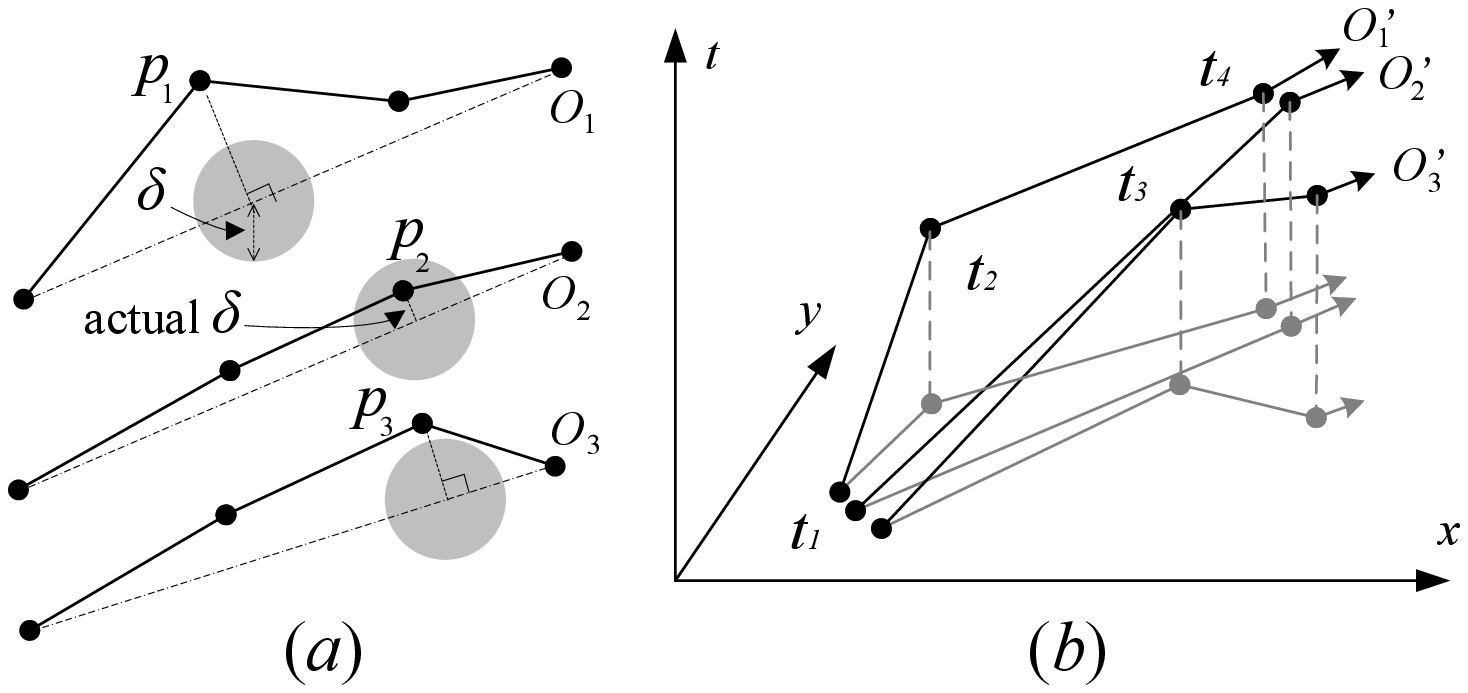}
    \caption{Trajectory Simplification}
    \label{fig_simple}
\end{figure}

Figure~\ref{fig_simple}(\emph{a}) illustrates the application of DP on the trajectories in
Figure~\ref{fig_convoy}. For $o_1$ trajectory, we first construct the virtual line between its end
points. Since the distance between the farthest point (i.e., $p_1$) and the virtual line exceeds
$\delta$, point $p_1$ will be kept in $o_1$'s corresponding simplified trajectory $o'_1$. Regarding
$o_2$, the distance of the furthest point (i.e., $p_2$) from the virtual line is below $\delta$;
thus, all intermediate points are removed from $o_2$'s simplified trajectory.
Figure~\ref{fig_simple}(\emph{b}) visualizes the simplified trajectories. Notice that each point in
a simplified trajectory corresponds to a point in the original trajectory and is associated with a
time value.

\vspace{0.2cm} \noindent \textbf{Measuring actual tolerances of simplified trajectories :}
We observe that an actual tolerance smaller than $\delta$ may exist so that the simplified
trajectory is valid. In the example of Figure~\ref{fig_simple}(\emph{a}), the actual tolerance of
$o'_2$ is determined by the distance between $p_2$ and the virtual line. We formally define the
actual tolerance as follows:

\begin{mydefinition} \label{def:act-tol}
\textbf{(Actual Tolerance)} Let $l'$ be a line segment in the
simplified trajectory $o'$, whose original trajectory is $o$.
The actual tolerance $\delta(l')$ of $l'$ is defined as:
$max_{t \in l'.\tau} \; D_{PL}(o(t),l')$.
The actual tolerance $\delta(o')$ of $o'$ is defined as the maximum
$\delta(l')$ value over all its line segments.
\end{mydefinition}

The actual tolerance of each line segment $l'$ of $o'$ can be computed
easily by examining the locations of $o$ during the corresponding time
interval $l'.\tau$.  In addition, the derivation of these tolerance
values can be seamlessly integrated into the DP algorithm so that the
original trajectory $o$ needs not be examined again.


The actual tolerances are valuable in the sense that they can be exploited to tighten the distance
computation for simplified trajectories, as we will show in the next section.

\subsection{Distance Bounds for Range Search} \label{sec:errorbound}

A simplified trajectory $o'$ may contain many omitted locations in comparison to its original
trajectory $o$. Thus, it is not possible to perform (density--connected) clustering at individual
time. If we generate virtual positions for the omitted points as done in CMC, there is no use for
the trajectory simplification.
The main challenge becomes one of performing clustering on the line segments of simplified
trajectories so that each snapshot cluster (on the original trajectories) is captured by a cluster
of line segments (from the simplified trajectories).

In density-based clustering techniques (e.g., DBSCAN), the core operation is $e$-neighborhood
search, i.e., to find objects within distance $e$ of a given object, at a fixed time $t$. We
proceed to develop the implementation of this core operation in the context of line segments. Let a
line segment $l'_q$ be given; our goal is then to retrieve all line segments $l'_i$ whose original
trajectory $o_i$ can possibly satisfy the condition $D(o_q(t),o_i(t)) \le e$ for some time point
$t$. This way, all qualifying convoy candidates are guaranteed to be found in the filter step.

\begin{figure}[hbt]
    \centering
    \includegraphics[width=5.5cm]{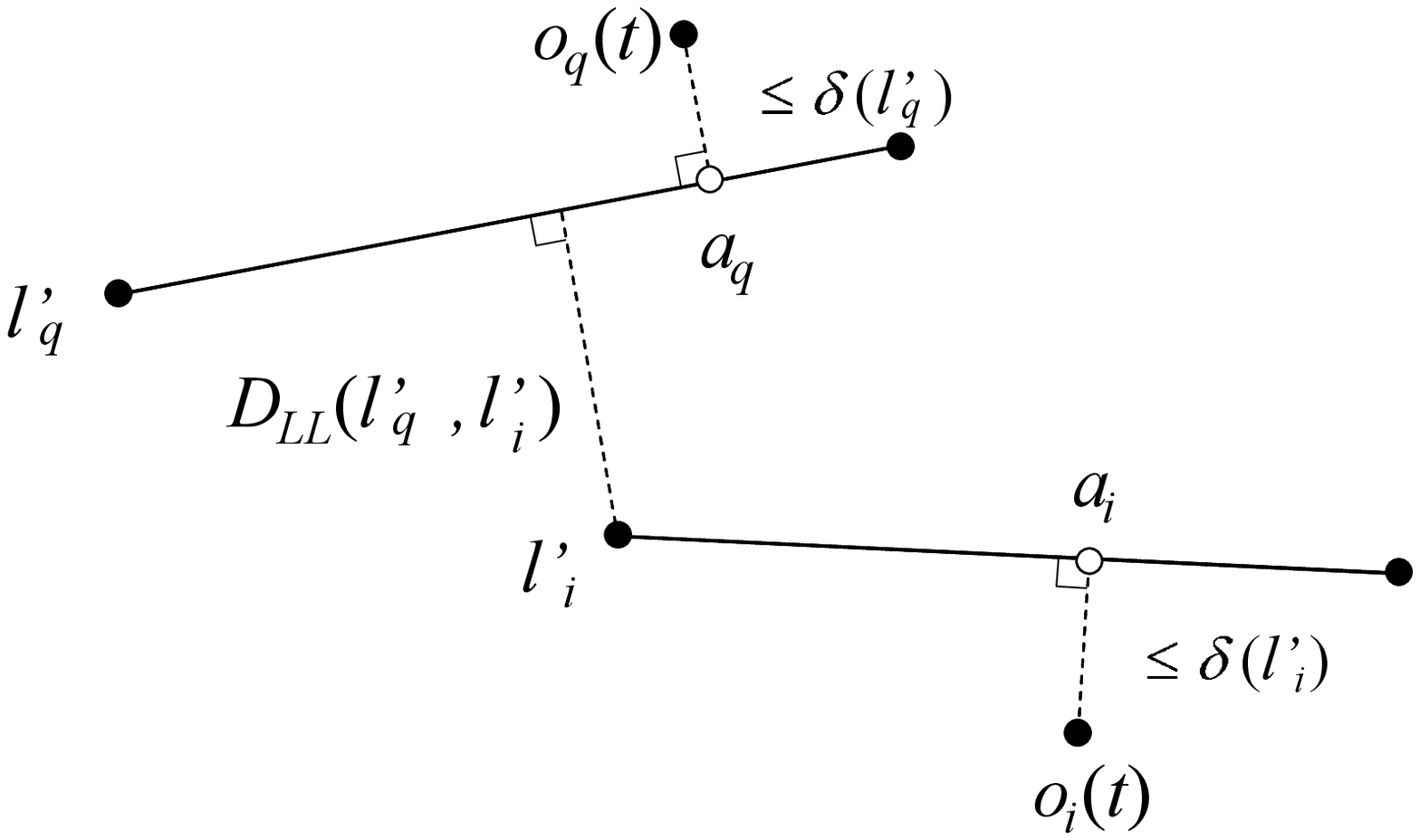}
    \caption{Trajectory Segments with Time Intervals Covering $t$}
    \label{fig_proof}
\end{figure}

Let $o'_q$ and $o'_i$ be simplified trajectories of the original trajectories $o_q$ and $o_i$. At a
given time $t$, the locations of $o_q$ and $o_i$ are $o_q(t)$ and $o_i(t)$. Observe that the
endpoints of line segments in $o'_q$ are timestamped. Let $l'_q$ be a line segment in $o'_q$ such
that its time interval $l'_q.\tau$ covers $t$. Similarly, we use $l'_i$ to denote the line segment
in $o'_i$ satisfying $t \in l'_i.\tau$. Figure~\ref{fig_proof} shows an example of two line
segments $l'_q$ and $l'_i$.

Lemma~\ref{lmm:prune-range} establishes the relationship between distances in the original
trajectories and those in the simplified trajectories.
\begin{lemma} \label{lmm:prune-range} Let $o'_q$ ($o'_i$) be the
  simplified trajectory of original trajectory $o_q$ ($o_i$). Given a
  time $t$, let $l'_q$ ($l'_i$) be the line segment in $o'_q$ ($o'_i$)
  with a time interval that covers $t$.

  If $D_{LL}(l'_q,l'_i) > e + \delta(l'_q) + \delta(l'_i)$ then
  $D(o_q(t),o_i(t)) > e$.
\end{lemma}
%
%
%
%

\begin{figure}[hbt]
    \centering
    \includegraphics[width=6.6cm]{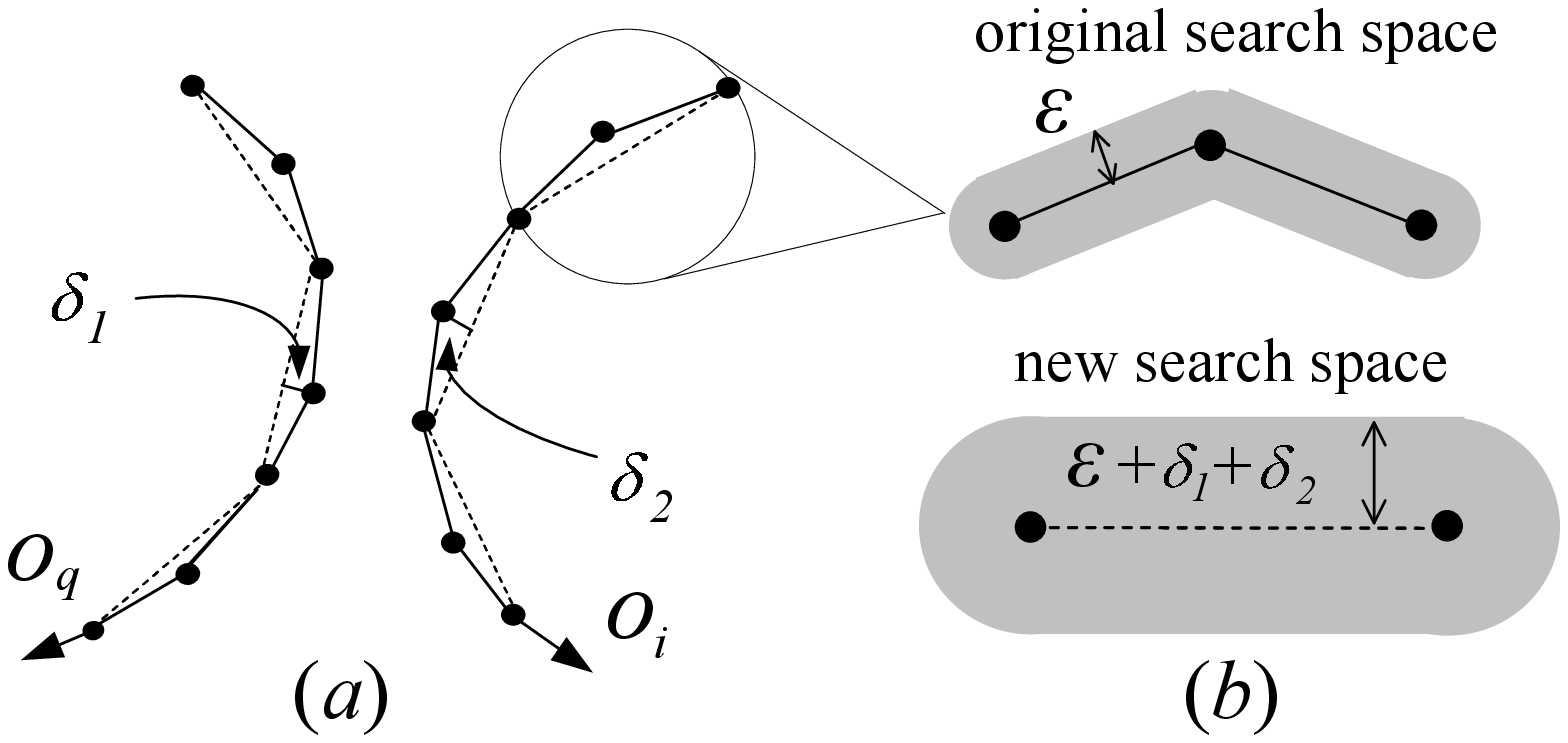}
    \caption{Range Search with Error Bounds}
    \label{fig_search}
\end{figure}

Lemma~\ref{lmm:prune-range} allows us to prune line segments $l'_i$ during the range search of the
given line segment $l'_q$. Figure~\ref{fig_search} illustrates the extended range for search over
simplified line segments with error bounds. In Figure~\ref{fig_search}(\emph{a}), half of the
points on the original trajectory are omitted (i.e., a 50\% reduction) with the given $\delta$
value. To enable correct discovery processing over the simplified trajectories (dotted lines), we
enlarge the search space as shown in Figure~\ref{fig_search}(\emph{b}).

\begin{figure}[hbt]
    \centering
    \includegraphics[width=8.5cm]{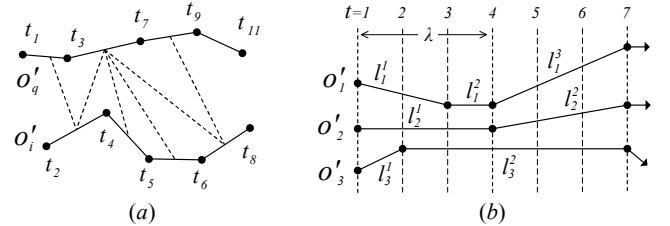}
    \caption{Measure of $\omega(o'_q,o'_i)$ and Time Partitioning}
    \label{fig_part}
\end{figure}

Notice that we still need to scan all $l'_i$ whose time intervals intersect with that of $l'_q$.
For example, the time interval [$t_3$,$t_7$] of the second line segment of $o'_q$ in
Figure~\ref{fig_part}(\emph{a}) intersects all of $o'_i$'s line segments. To obtain better
performance, we intend to prune a subset $S$ of line segments fast. During the range search of the
given line segment $l'_q$, Lemma~\ref{lmm:prune-group-range}, next, enables us to prune an
non-qualifying $S$ before examining its line segments. The proofs of Lemma~\ref{lmm:prune-range}
and Lemma~\ref{lmm:prune-group-range} are provided in the appendix.

\begin{lemma} \label{lmm:prune-group-range} Let $S$ be a subset of
line segments $l'_i$ (from simplified trajectories). Let $\mathcal{B}(S)$ be the minimum bounding
box of all segments in $S$, $S.\tau = \bigcup_{l'_i \in S} l'_i.\tau$, and $\delta_{max}(S) =
\max_{l'_i \in S} \delta(l'_i)$. Let line segment $l'_q$ have a time interval that intersects with
that of $S$, i.e., $S.\tau \cap l'_q.\tau \ne \emptyset$.

If $D_{min}(\mathcal{B}(l'_q),\mathcal{B}(S)) > e + \delta(l'_q) +
\delta_{max}(S)$ then \\
$D(o_q(t),o_i(t)) > e$ holds for all $l'_i \in S$.
\end{lemma}
%
%

%
We proceed to outline how to perform range search for $l'_q$ in multiple steps by gradually
tightening the condition: First, we retrieve a set of line segments $S$ whose time intervals
overlap with that of $l'_q$. We then apply Lemma~\ref{lmm:prune-group-range} to prune
non-qualifying line segments in $S$ at an early stage. Next, for each remaining line segment in
$S$, we discard non-qualifying line segments by applying Lemma~\ref{lmm:prune-range}. Any surviving
line segment is included in the $e$-neighborhood of the line segment $l'_q$. Using this multi-step
range search for line segments, we are able to perform density--connected clustering of line
segments efficiently.


\noindent \textbf{Extension for trajectories :}
So far, we have addressed range search only for line segments. In fact, it is feasible to
generalize the search to apply to an entire trajectory. And by applying clustering on trajectories
directly, we further reduce the cost of the filter step. As we will see in the next section, the
technique below is applicable to sub-trajectories as well, enabling us to control the granularity
of the filter step.

We aim to retrieve all simplified trajectories $o'_i$ whose original trajectories $o_i$ possibly
satisfy the condition $D(o_q(t),o_i(t)) \le e$ for some time $t$. In case $o'_q$ and $o'_i$ have
disjoint time intervals (i.e., $o'_q.\tau \cap o'_i.\tau = \emptyset$), they cannot belong to the
same convoy. Otherwise, we define their $\omega$ value as follows:

$\omega(o'_q,o'_i)= \min \{ D_{LL}(l'_q,l'_i) - \delta(l'_q) - \delta(l'_i) \; | \; l'_i \in o'_i,
\; l'_q \in o'_q, $
\\*
$l'_q.\tau \cap l'_i.\tau \ne \emptyset\}$.

Figure~\ref{fig_part}(\emph{a}) shows an example of computing the $\omega(o'_q,o'_i)$ value between
two simplified trajectories $o'_q$ and $o'_i$. Line segments with shared time interval are linked
by dotted lines, contributing a term in the value of $\omega(o'_q,o'_i)$. 
%
If $\omega(o'_q,o'_i) > e$, no time $t$ exists such that
$D(o_q(t),o_i(t)) \le e$. Otherwise, their locations in the original
trajectories may be within distance $e$ for some time $t$.

\subsection{The CuTS Algorithm} \label{sec:adaptive}

We first present a general overview of the CuTS (\underline{C}onvoy Discovery \underline{u}sing
\underline{T}rajectory \underline{S}implification) algorithm, then illustrate aspects of the
algorithm with examples, and finally present the details of the algorithm.

In the filter step, we first apply simplification (with tolerance $\delta$) to the original
trajectories in order to obtain their simplified trajectories. We then partition the time domain
(with each partition covering $\lambda$ time points) and assign the line segments of each $o'_i$ to
qualifying partitions. Next, we perform clustering on those line segments. Clusters across adjacent
partitions with common objects are used to form convoy candidates. In the refinement step, we
perform clustering of the original trajectories of the objects in each convoy candidate. The total
computational cost of the CuTS algorithm is the sum of the simplification, the clustering, and the
refinement costs. Our experiments in Section~\ref{sec:exp} suggest that the simplification and
refinement costs are very low in practice.

To understand the filter step of CuTS better, consider Figure~\ref{fig_part}(\emph{b}) where the
time domain is divided into equal-length ($\lambda=4$) partitions $\mathcal{T}_1$ and
$\mathcal{T}_2$ with time intervals $[t_1,t_4]$ and $[t_4,t_7]$, respectively.
%
%
The time partition $\mathcal{T}_1$ contains the following line
segments: $l^1_1$ and $l^2_1$ of $o'_1$, $l^1_2$ of $o'_2$, and
$l^1_3$ and $l^2_3$ of $o'_3$.
%
%
Note that the line segment $l^2_3$ will be inserted into both $\mathcal{T}_1$ and $\mathcal{T}_2$
to avoid any possible false dismissal when we compute the value of $\omega(o'_q,o'_i)$ in
Figure~\ref{fig_part}(\emph{a}).


\vspace{0.2cm} \noindent \textbf{Algorithm description.}
Algorithm~\ref{alg:CuTS_filter} presents the pseudocode of CuTS's filter step. In addition to the
convoy query parameters $m$, $k$, and $e$, two internal parameters $\delta$ (tolerance for
trajectory simplification) and $\lambda$ (the length of each partition) also need to be specified.
Those parameter values are relevant to the performance only (e.g., execution time) and do not
affect the correctness. Guidelines for choosing their values will be presented in
Section~\ref{sec:param}.


\begin{algorithm}[hbt]
\small
\caption{\bf CuTS\_Filter (Object set $O$, Integer $m$, Integer $k$,
Distance threshold $e$)}
\begin{algorithmic}[1]
    \STATE $\delta \leftarrow$ ComputeDelta($O$, $e$)
    \FOR {each trajectory $o_i \in O$}
        \STATE $o'_i \leftarrow$ Douglas-Peucker($o_i, \delta$)
    \ENDFOR
    \STATE $\lambda \leftarrow$ ComputeLambda($O$, $k$, $(\sum_i |o_i|  ) / (\sum_i |o'_i| )$)
    \STATE $V \leftarrow \emptyset$ 
    \STATE divide the time domain into $\lambda$-length disjoint partitions
    \FOR {each time partition $\mathcal{T}_z$ (in ascending order)}
        \STATE $V_{next} \leftarrow \emptyset$
        \FOR {each $o'_i$ satisfying $o'_i.\tau \cap \mathcal{T}_z.\tau \ne \emptyset$}
            \STATE insert $l^j_i \in o'_i$ (intersecting time interval of $\mathcal{T}_z$) into $\mathcal{G}$
        \ENDFOR
        \STATE $C \leftarrow$ TRAJ-DBSCAN($\mathcal{G}, e, m$)
        \FOR {each convoy candidate $v \in V$}
            \STATE $v.$assigned $\leftarrow$ false
            \FOR {each cluster $c \in C$}
                \IF {$|c \cap v| \geq m$}
                    \STATE $v$.assigned $\leftarrow$ true
                    \STATE $v' \leftarrow c \cap v$
                    \STATE $v'$.lifetime $\leftarrow v$.lifetime + $\lambda$
                    \STATE $V_{next} \leftarrow V_{next} \cup v'$
                    \STATE $c$.assigned $\leftarrow$ true
                \ENDIF
            \ENDFOR
            \IF {$v$.assigned = false and $v$.lifetime$\ge k$}
                \STATE  $V_{cand} \leftarrow V_{cand} \cup v$
            \ENDIF
        \ENDFOR
        \FOR {each $c \in C$}
            \IF {$c$.assigned = false} 
                \STATE $c$.lifetime $\leftarrow \lambda$
                \STATE $V_{next} \leftarrow V_{next} \cup c$
            \ENDIF
        \ENDFOR
        \STATE $V \leftarrow V_{next}$
     \ENDFOR
     \RETURN $V_{cand}$
\end{algorithmic}
\label{alg:CuTS_filter}
\end{algorithm}

Lines~2--3 of the algorithm perform trajectory simplification for all objects. Next, the time
domain is partitioned, each partition holding $\lambda$ consecutive time points.
%
%
Time partitions are then processed iteratively in ascending order of their time. Let the current
loop consider the time partition $\mathcal{T}_z$. The algorithm builds a polyline (i.e., a sequence
of line segments) from a simplified trajectory $o'_i$, which contains the line segments of $o'_i$
whose time intervals intersect to $\mathcal{T}_z$. It then stores all the polylines from each
simplified trajectory into a data structure $\mathcal{G}$. Next, density clustering is performed
for the sub-trajectories in $\mathcal{G}$ (see Line~11).

The set $V$ keeps track of the convoy candidates found in previous iterations, whereas the set
$V_{next}$ stores new candidates found in the current iteration. For Lines~12--20, each cluster $c
\in C$ (found in the current iteration) is joined with those in $V$, as long as their intersections
have at least $m$ objects. Also, candidate convoys with lifetime above $k$ are inserted into the
candidate set $V_{cand}$. Clusters that cannot join with previous convoy candidates are then
considered as new candidates (Lines~23--26).

Finally, Algorithm~\ref{alg:CuTS_verify} contains the pseudocode of the refinement step of the CuTS
algorithm. Suppose that $v$ is the convoy candidate in the candidate set $V$ that is currently
being examined. We first determine the time interval $[t_{start}, t_{end}]$ for $v$ and then
identify the set $O'$ of the original trajectories whose line segments appear in $v$. Finally, we
apply CMC for trajectories in $O'$, considering only time points in the interval $[t_{start},
t_{end}]$.

\begin{algorithm}[hbt]
\small
\caption{\bf CuTS\_Refinement (Candidate set $V_{cand}$, Object set
$O$, Integer $m$, Integer $k$, Distance threshold $e$)}
\begin{algorithmic}[1]
    \FOR {each $v \in V_{cand}$}
        \STATE $t_{start} \leftarrow$ start time of $v$
        \STATE $t_{end} \leftarrow$ end time of $v$
        \STATE $O' \leftarrow \{ o_i \in O  \; | \; l^j_i.\tau \in v \cap l^j_i.\tau \in o_i \}$
        \STATE call CMC($O', m, k, e$) with the time interval $[t_{start},t_{end}]$
    \ENDFOR
\end{algorithmic}
\label{alg:CuTS_verify}
\end{algorithm}

\section{Extensions of CuTS} \label{sec:ext}

In this section, we introduce two enhancements of CuTS. One accelerates the process of trajectory
simplification and brings higher efficiency. The other shortens the search range for clustering by
considering temporal information of trajectories, reducing the number of candidates after the
filter step of CuTS.

\subsection{Faster Trajectory Simplification - CuTS+} \label{sec:cuts+}

The Douglas-Peucker algorithm (DP) utilizes the divide-conquer technique (see
Section~\ref{sec:trjsimple}). It is well-known that techniques built on the divide-conquer paradigm
show the best performance if a given input is divided into two sub-inputs equally in each division
step. Inspired by this, we modify the original DP algorithm for speeding up the simplification
process, obtaining DP+.

\begin{figure}[!h]
    \centering
    \includegraphics[width=8.5cm]{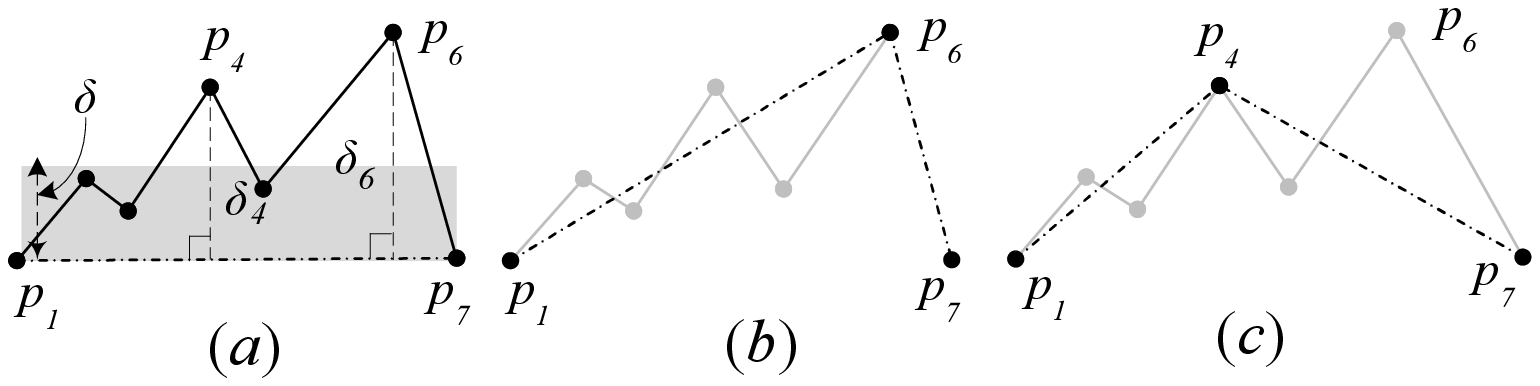}
    \caption{Comparison between DP and DP+}
    \label{fig_dp+}
\end{figure}

Specifically, DP+ selects the closest point to the middle of a given trajectory among the points
exceeding tolerance value $\delta$ at each approximation step. Figure \ref{fig_dp+}(\emph{a})
demonstrates an original trajectory having seven points, which has two intermediate points $p_4$
and $p_6$ whose distances from $\overline{p_1p_7}$ are greater than the given $\delta$ value (the
gray area in the figure). The DP method selects the point having the largest distance (i.e.,
$\delta_6$); hence, the result of this division step will be as shown in Figure
\ref{fig_dp+}(\emph{b}).

In contrast, our DP+ method picks the point $p_4$ that is the closest to the middle point of ${p_1,
p_2, \cdots, p_7}$ among intermediate points exceeding $\delta$ (i.e., $p_4$ and $p_6$). This
technique divides $\overline{p_1p_7}$ into two sub-trajectories $\overline{p_1p_4}$ and
$\overline{p_4p_7}$, which have similar numbers of points (Figure \ref{fig_dp+}(\emph{c})).
Therefore, the whole process of trajectory simplification is expected to be more efficient.

\pagebreak

Compared with DP, DP+ may have lower simplification power. In fact, each division process of DP+
does not preserve the shape of a given original trajectory well; hence, the next division process
may not be as effective as that of DP. For example, in Figure~\ref{fig_dp+}(\emph{c}), $p_6$ will
be kept using DP+ because $D_{PL}(p_6,\overline {p_4 p_7}) > \delta$, and then the simplified
trajectory will be $p_1,p_4,p_6,p_7$, whereas $p_1,p_6,p_7$ will be the result of DP in
Figure~\ref{fig_dp+}(\emph{b}).

In spite of the lower reduction, DP+ can enhance the discovery processing of CuTS in two areas.
First, note that we are interested in efficient discovery of convoys in this study. As long as the
search distances are bounded, faster simplification of trajectories can play a more important role
in finding convoys. Second, the actual tolerances obtained by DP+ are always smaller or equal to
those obtained by DP (e.g., $\delta_4 < \delta_6$ in the example). This tightens the error bounds
of range search for clustering, leading a more effective filter step.

We extend CuTS to CuTS+, which is built on the DP+ simplification method. All other discovery
processes of CuTS+ are the same as those of CuTS.

\subsection{Temporal Extension - CuTS*} \label{sec:cuts*}

Recall that CuTS applies trajectory simplification (DP) on original trajectories in the filter
step. However, as we will see shortly, intermediate locations on simplified line segments cannot be
associated with fixed timestamps. Consequently, the bounds on distances between line segments may
not be tight, the result being that overly many convoy candidates can be produced in the filter
step. This may yield a more expensive refinement step.

In this section, we extend CuTS to CuTS* by considering temporal aspects for both the trajectory
simplification and the distance measure on simplified trajectories. This enables us to tighten
distance bounds between simplified trajectories, improves the effectiveness of the filter step.

%



\vspace{0.3cm} \noindent \textbf{Comparison between DP and DP*}: We discussed the differences
between the two trajectory simplification techniques DP \cite{DP} and DP* \cite{st-simple1} in
Section~\ref{sec:trjsimple}. In Figure~\ref{fig_stsimpleFT}(\emph{b}), DP* translates the time
ratio of $p_2$ between $p_1$ and $p_3$ into a location $p'_2$ on the line segment
$\overline{p_1p_3}$. Since $p_2$ exceeds the $\delta$ range of $p'_2$, the point $p_2$ is kept in
the simplified trajectory $o'_1$, which is different from DP.

From the example, we can see that DP* has a lower vertex reduction ratio for trajectories.
Nevertheless, DP* permits us to derive tighter distance measures between trajectory segments,
improving the overall effectiveness of the filter step.

\begin{figure}[hbt]
    \centering
    \includegraphics[width=7.2cm]{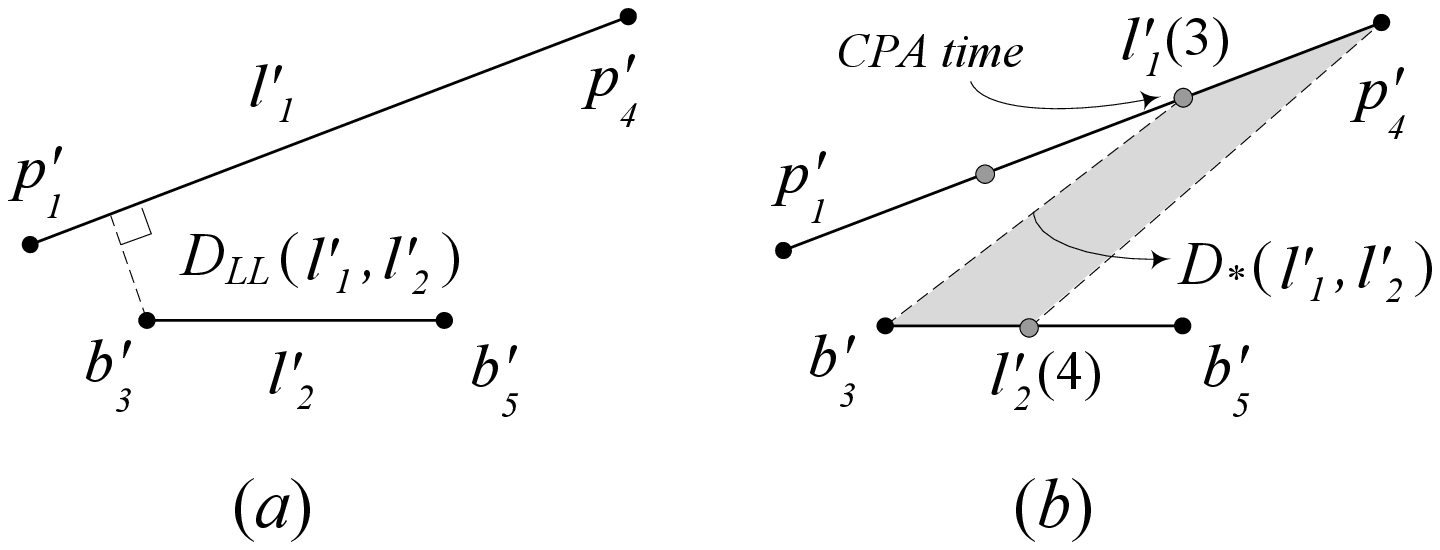}
    \caption{Different Distance Measures of Trajectory Segments}
    \label{fig_dist3d}
\end{figure}

\pagebreak

Figure~\ref{fig_dist3d}(\emph{a}) shows two simplified line segments $l'_1$ and $l'_2$, obtained
from DP. Here, $l'_1$ has the endpoints $p'_1$ and $p'_4$, corresponding to its locations at times
$t_1$ and $t_4$. Similarly, $l'_2$ has endpoints $b'_3$ and $b'_5$, corresponding to its locations
at times $t_3$ and $t_5$. The shortest distance between $l'_1$ and $l'_2$ is given by
$D_{LL}(l'_1,l'_2)$.

Figure~\ref{fig_dist3d}(\emph{b}) contains simplified line segments from DP*. Since DP* captures
the time ratio in the simplified line segment, we are able to derive the locations $l'_1(3)$ on
$l'_1$ and $l'_2(4)$ on $l'_2$. Let $l'_p=\{p_u,p_v\}$ be a simplified line segment having a time
interval $l'_p.\tau=[u,v]$. The location of $l'_p$ at a time $t \in [u,v]$ is defined as:
$$l'_p(t) = p_u + \frac{t-u}{v-u} (p_v-p_u)$$
Note that the terms $l'_p(t)$, $p_u$, and $(p_v-p_u)$ are 2D vectors representing locations.

Before defining $D_*(l'_1,l'_2)$ formally, we need to introduce the time of the \emph{Closest Point
of Approach}, called the CPA time ($t_{CPA}$) \cite{st-join6}. This is the time when the distance
between two dynamic objects is the shortest, considering their velocities. Let $l'_q=\{q_w,q_x\}$
be another simplified line segment during $l'_q.\tau=[w,x]$. The CPA time of $l'_p$ and $l'_q$ is
computed by :
$$t_{CPA}= \frac{-(p_u - q_w) \cdot (l'_p(t) - l'_q(t))}{|l'_p(t) - l'_q(t)|^2}$$
where, $l'_q(t)$, $q_w$, and $(q_w-q_x)$ are also location vectors.

Observe that the common interval of $l'_1$ and $l'_2$ is $[t_3, t_4]$ (gray area in
Figure~\ref{fig_dist3d}(\emph{b})). The {\em tightened} shortest distance $D_*(l'_1,l'_2)$ between
them is computed as :
$$D_*(l'_1,l'_2)= D(l'_1(t_{CPA}),l'_2(t_{CPA}))\hspace{0.5cm} t_{CPA} \in (l'_1.\tau \cap l'_2.\tau)$$

When their time intervals do not intersect, i.e., $l'_1.\tau \cap l'_2.\tau = \emptyset$, their
distance is set to $\infty$ .

Clearly, $D_*(l'_1,l'_2)$ is longer than $D_{LL}(l'_1,l'_2)$; hence, the line segments in
Figure~\ref{fig_dist3d}(\emph{b}) have a lower probability of forming a cluster together than do
those in Figure~\ref{fig_dist3d}(\emph{a}). These tightened distance bounds improve the
effectiveness of the filter step.


\vspace{0.2cm} \noindent \textbf{Distance bounds for DP* simplified line segments}:
Using the notations from Lemma~\ref{lmm:prune-range}, we derive the counterpart that uses the
tightened distance $D_*$ between line segments (as opposed to the distance $D_{LL}$).
Lemma~\ref{lmm:prune-star-range} establishes the relationship between distances in original
trajectories and those in simplified trajectories (obtained by DP*). The proof is provided in the
appendix.

\begin{lemma} \label{lmm:prune-star-range} Suppose that $o'_q$
  ($o'_i$) is the simplified trajectory (from DP*) of the original
  trajectory $o_q$ ($o_i$). Given a time $t$, let $l'_q$ ($l'_i$)
  be the line segment in $o'_q$ ($o'_i$) with time interval covering
  $t$.

  If $D_{*}(l'_q,l'_i) > e + \delta(l'_q) + \delta(l'_i)$ then
  $D(o_q(t),o_i(t)) > e$.
\end{lemma}
%


\noindent \textbf{CuTS* algorithm for convoy discovery}:
We develop an enhanced algorithm, called CuTS*, to exploit the above tightened distance bounds for
query processing. Two components of CuTS need to be replaced. First, CuTS* applies DP* for the
trajectory simplification. Second, during density clustering in the filter step,
Lemma~\ref{lmm:prune-star-range} is utilized in the range search operations (as opposed to
Lemma~\ref{lmm:prune-range}). The above modifications improve the effectiveness of the filter step
in CuTS*. The following table summarizes the key components of CuTS and its extensions.

\begin{table}[hbt]
\centering
\begin{tabular}{|c|c|c|c|c|}
\hline {\bf Method}          & {\bf CuTS}  & {\bf CuTS+}  & {\bf CuTS*}  \\
\hline  simplification   &  DP \cite{DP}   & DP+ [Section~\ref{sec:cuts+}]          & DP* \cite{st-simple1} \\
\hline distance function     &  $D_{LL}$       & $D_{LL}$        & $D_*$ \\
\hline
\end{tabular}
\end{table}

\section{Experiments} \label{sec:exp}


In this experimental study, we first compare the discovery efficiency between CMC, which is an
adaption of a moving-clustering algorithm (MC2) \cite{st-clustering1} for our convoy discovery
problem, and the CuTS family (CuTS, CuTS+, and CuTS*). We then analyze the performance of each
method of the CuTS family while varying the settings of their key parameters.

We implemented the above algorithms in the C++ language on a Windows Server 2003 operating system.
The experiments were performed using an Intel Xeon CPU 2.50 GHz system with 16GB of main memory.


\subsection{Dataset and Parameter Setting}

For studying the performance of our methods in a real-world setting, we used several real datasets
that were obtained from vehicles and animals. Due to the different object types, their trajectories
have distinct characteristics, such as the frequency of location sampling and data distributions.
The details of each dataset are described as follows:

\vspace{0.2cm}

\vspace{0.1cm} \noindent \textbf{Truck}: We obtained 276 trajectories of 50 trucks moving in the
Athens metropolitan area in Greece \cite{rtreeportal}. The trucks were carrying concrete to several
construction sites for 33 days while their locations were measured. To be able to find more
convoys, we regarded each trajectory as a distinct truck's trajectory and removed the day
information from the data. Thus, the dataset became 276 trucks' movements on the same day.

\vspace{0.1cm} \noindent \textbf{Cattle}: To reduce a major cost for cattle producers, a virtual
fencing project in CSIRO, Australia studied managing herds of cattle with virtual boundaries. We
obtained 13 cattle's movements for several hours from the project. Their locations were provided by
GPS-enabled ear--tags every second. A distinguishable aspect of this dataset is its very large
number of timestamps.

\vspace{0.1cm} \noindent \textbf{Car}: Normal travel patterns of over 500 private cars were
analyzed for building reasonable road pricing schemes in Copenhagen, Denmark. We obtained 183 cars'
trajectories during one week \cite{daisy}. Trajectories in this dataset had very different lengths.

\vspace{0.1cm} \noindent \textbf{Taxi}: The GPS logs of 500 taxis in Beijing, China were recorded
during a day and studied in Institute of Software, Chinese Academy of Sciences. The locations of
the trajectories were sampled irregularly. For example, some taxis reported their locations every
three minutes, while some did it once in several minutes.

\vspace{0.2cm}

%

In our experiments, we defined a convoy as containing at least 3 objects (except Cattle due to the
small number of objects) that travel closely for 3 minutes (i.e., $m=3$ and $k=180$). We also
adjusted the values of neighborhood range $e$ to be able to find 1 to 100 convoys for each dataset.
To perform convoy discovery using our main methods (CuTS, CuTS+, and CuTS*), we still need to
determine two key parameters, namely the tolerance value ($\delta$) for trajectory simplification
and the length of time partition ($\lambda$). These parameter values were computed by our
guidelines that will be discussed in Section~\ref{sec:param}.

Table~\ref{tab:data} provides (i) detailed information of each dataset,
 (ii) the settings of the parameters to be used throughout our experiments, and (iii) the number
of convoys discovered by our proposed methods with the parameters.

\begin{table}[hbt]
\centering
\small
\begin{tabular}{|c|c|c|c|c|}
\hline                                  & {\bf Truck}  & {\bf Cattle}  & {\bf Car} & {\bf Taxi}   \\
\hline number of objects ($N$)          &  267         & 13            & 183              & 500   \\
\hline time domain length ($T$)         &  10586       & 175636        & 8757             & 965   \\
\hline average trajectory length        &  224         & 175636        & 451              & 82   \\
\hline data size (points)               &  59894       & 2283268       & 82590            & 41144 \\
\hline\hline number of convoy objects ($m$) &  3           & 2             &  3               & 3   \\
\hline convoy lifetime ($k$)            &  180         & 180           & 180              & 180   \\
\hline neighborhood range ($e$)         &  8           & 300           & 80               & 40   \\
\hline simplification tolerance ($\delta$)&  5.9         & 274.2         & 63.4             & 31.5  \\
\hline time partition length ($\lambda$)&  4           & 36            & 24               & 4   \\
\hline\hline number of convoys discovered   &  91          & 47            & 15               & 4   \\

\hline
\end{tabular}
\caption{Settings for Experiments}\label{tab:data}
\end{table}

\subsection{CMC vs. The CuTS Family}\label{exp:cmc}

\begin{figure}[hbt]
    \center
    \includegraphics[width=8.5cm]{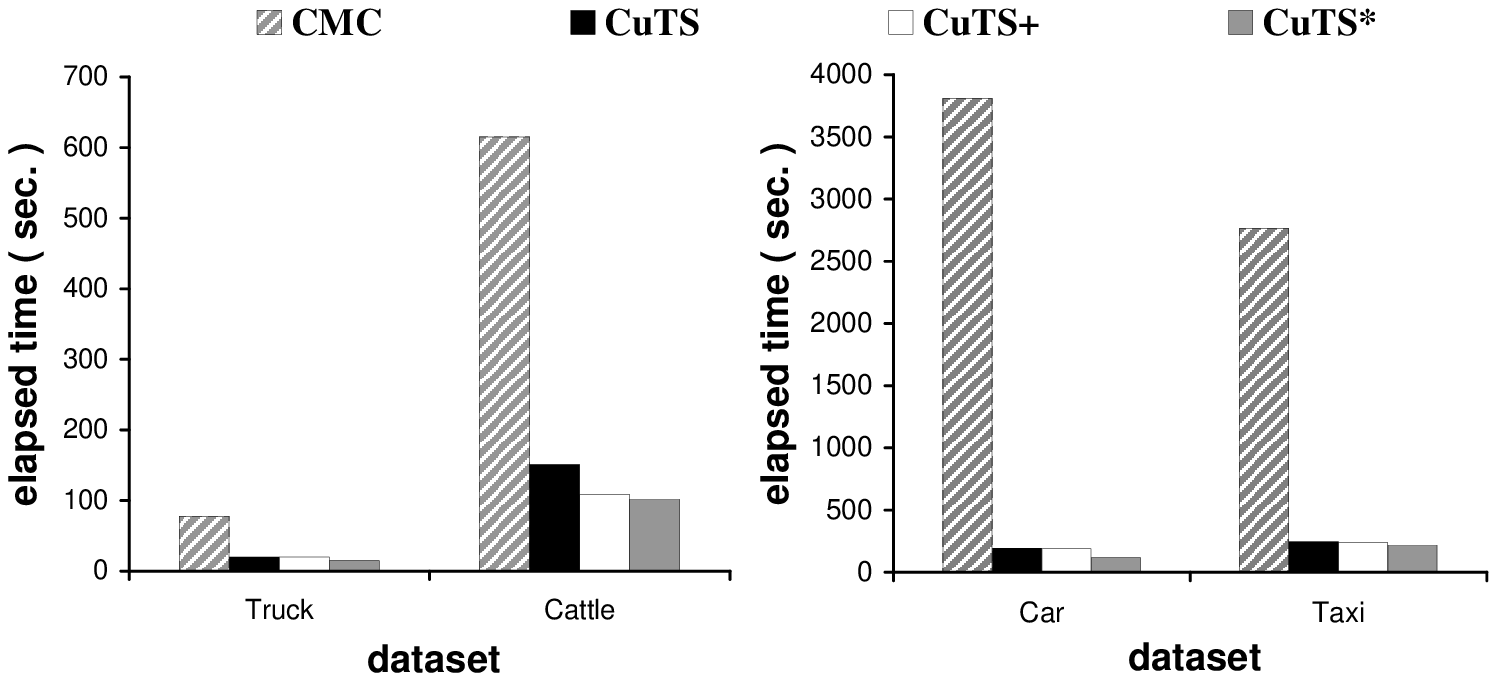}
    \caption{Comparisons of Query Processing Time}
    \label{exp_cmc}
\end{figure}

First, we compared the efficiency of CMC versus the CuTS family. Over all the datasets, the CuTS
family was 3.9 times (at least) to 33.1 times (at most) faster than CMC, as seen in
Figure~\ref{exp_cmc}, and especially CuTS* had the highest efficiency. The performance differences
were more obvious in the Car and the Taxi datasets though their data sizes (total number of points)
were less than 10\% of Cattle's data size. Since those two datasets had many numbers of missing
points and different lifetimes of each trajectory, CMC incurred extra computational cost to make
virtual points for those missing times to measure density-connection correctly (see
Section~\ref{sec:cmc}). It also caused a considerable growth of the actual data size for the
discovery processing. Notice that our main methods, the CuTS family, can perform the discovery
without any extra processing regardless of the number of missing points.


\begin{figure}[!h]
    \center
    \includegraphics[width=8.5cm]{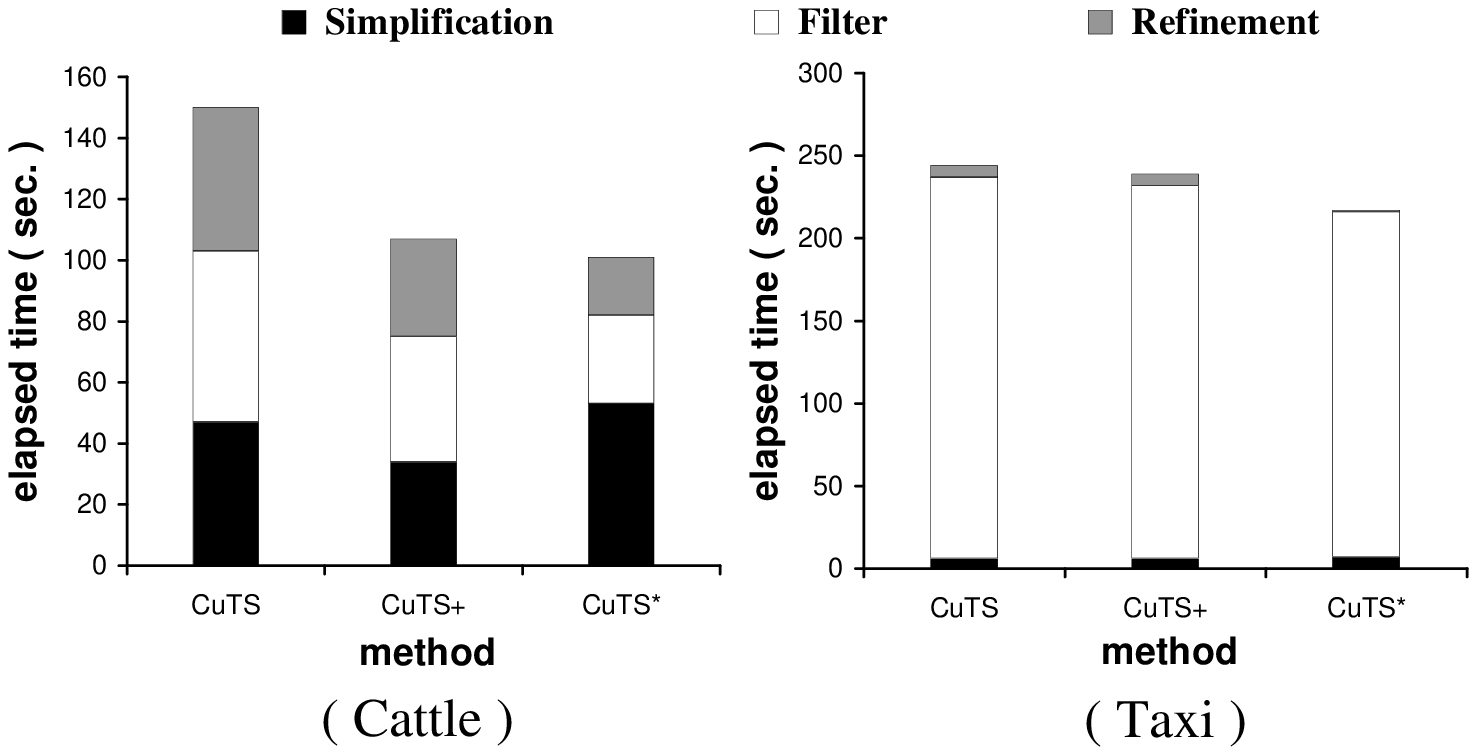}
    \caption{Analysis of Query Processing Cost}
    \label{exp_anal}
\end{figure}

In Figure~\ref{exp_anal}, we report on the elapsed times of each method of the CuTS family for the
Cattle and Taxi datasets (magnified views of the results in Figure~\ref{exp_cmc}). For brevity, we
show the two most distinctive results only. In the results for the Cattle dataset, the
simplification cost dominates for all the methods. In general, convoy processing is more sensitive
to the number of objects $N$ than to the number of timestamps $T$ since the clustering method
(DBSCAN) has $O(N^2)$ computational cost ($O(N \cdot \log N)$ with a spatial index). The Cattle
dataset has only 13 objects, and the cost of each clustering is very low though it is performed $T$
times. As a result, the total discovery times are more influenced by the simplification process
than the filtering and refinement steps.

\vspace{0.2cm}

The reason why each simplification method has different efficiency will be studied in
Section~\ref{exp:cuts}. Recall that, although the CuTS family needs much time for trajectory
simplification on the Cattle dataset, their total discovery times are still much lower than those
of CMC in the previous experiments.


Another interesting observation found with the Cattle data is that CuTS+ has not only faster
trajectory simplification, but also lower refinement cost. This is because DP+ as used in CuTS+ has
not only higher efficiency of simplification, but also tighter error bounds than DP as used by
CuTS, as described in Section~\ref{sec:cuts+}. 

Compared to the Cattle data, trajectory simplification had very low computational cost on the Taxi
dataset. As the Taxi dataset has a short $T$ but a larger $N$, the clustering cost dominates the
discovery time. In addition, since the number of convoy candidates was small for this data (will be
shown in the next experiments), only little refinement was necessary.

For the other two datasets, the composition of computational time was about 70\%-80\% for filtering
(around 5\%-15\% for trajectory simplification) and 20\%-30\% for refinement. Therefore, it is very
reasonable to `invest' some time in trajectory simplification.

We also studied the effect of using the actual tolerance for the range search of clustering. When
we perform the trajectory simplification, we use the tolerance value $\delta$, named the
\emph{global tolerance} here. The key process of the simplification is to remove intermediate
points whose distances from the virtual line linking two end points of the original trajectory do
not exceed $\delta$. Any distance of those removed points (i.e., \emph{actual tolerance}) is always
smaller than or equal to the global tolerance (see Section~\ref{sec:simple}). The actual tolerance
is useful for range search since the search area should be reduced.
\begin{figure}[hbt]
    \center
    \includegraphics[width=8.5cm]{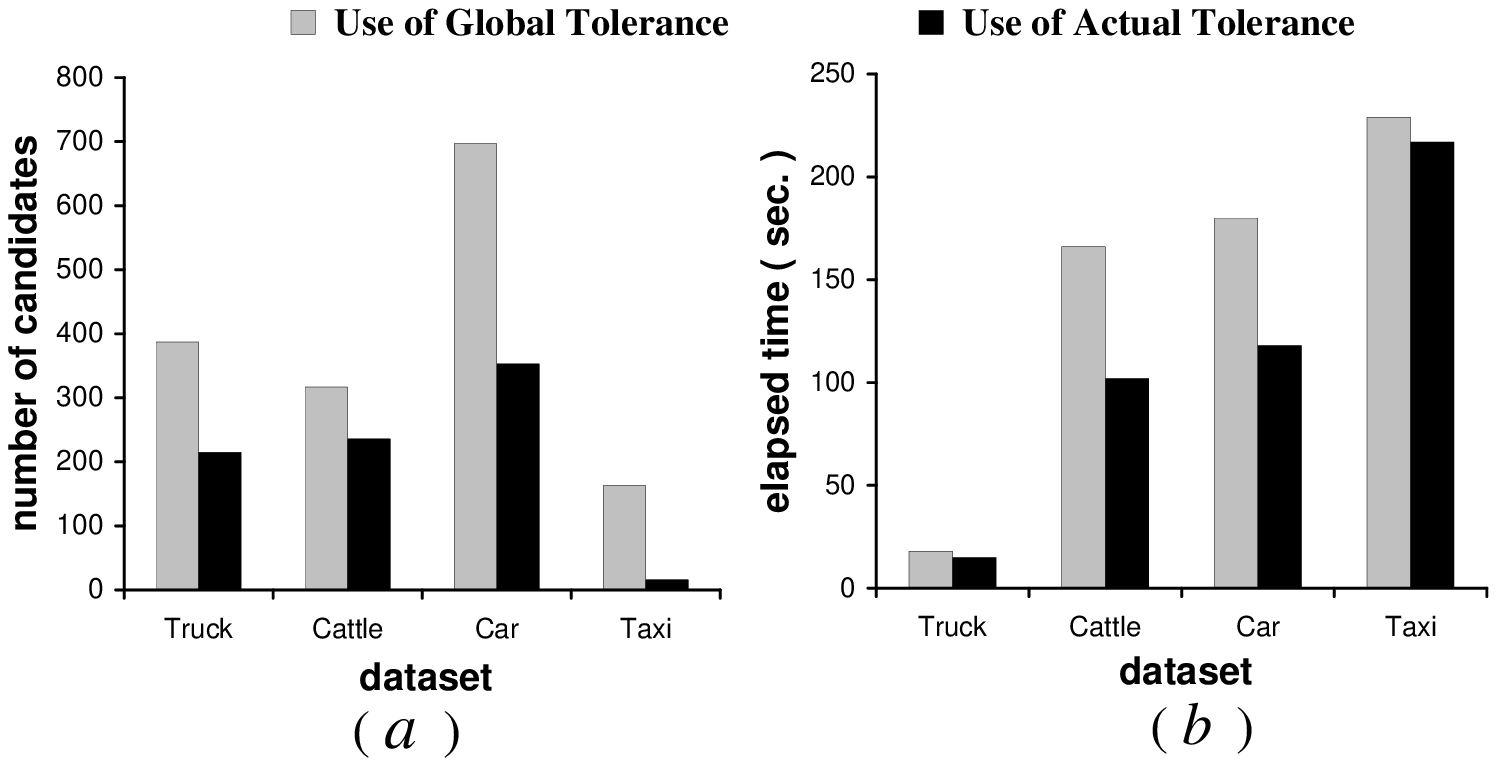}
    \caption{Effect of Actual Tolerance}
    \label{exp_acttol}
\end{figure}

Figure~\ref{exp_acttol}(\emph{a}) demonstrates the filtering power of the global and actual
tolerances for CuTS*. We omit the results for CuTS and CuTS+ because they are similar. As shown,
the number of candidates after the filtering step decreases considerably when we use the actual
tolerance. The advantage of the improved filtering by the actual tolerance is reflected in the
efficiency of convoy discovery as shown in Figure~\ref{exp_acttol}(\emph{b}). Yet, the effect is
relatively small on the Truck and the Taxi datasets. This is because some candidates that do not
need much computation for the refinement step are pruned when using the actual tolerance. We
present a more precise way of measuring the filter's effectiveness in the following section.

\subsection{CuTS vs. CuTS+ vs. CuTS*} \label{exp:cuts}

We have already discussed different techniques for trajectory simplification. The difference
between the original Douglas-Peucker algorithm (DP) and its temporal extension DP* was covered in
Section~\ref{sec:trjsimple}. We also developed a DP variant, named DP+, in Section \ref{sec:cuts+}.
It is of interest to compare the performance of those methods.

\begin{figure}[hbtt]
    \center
    \includegraphics[width=8.5cm]{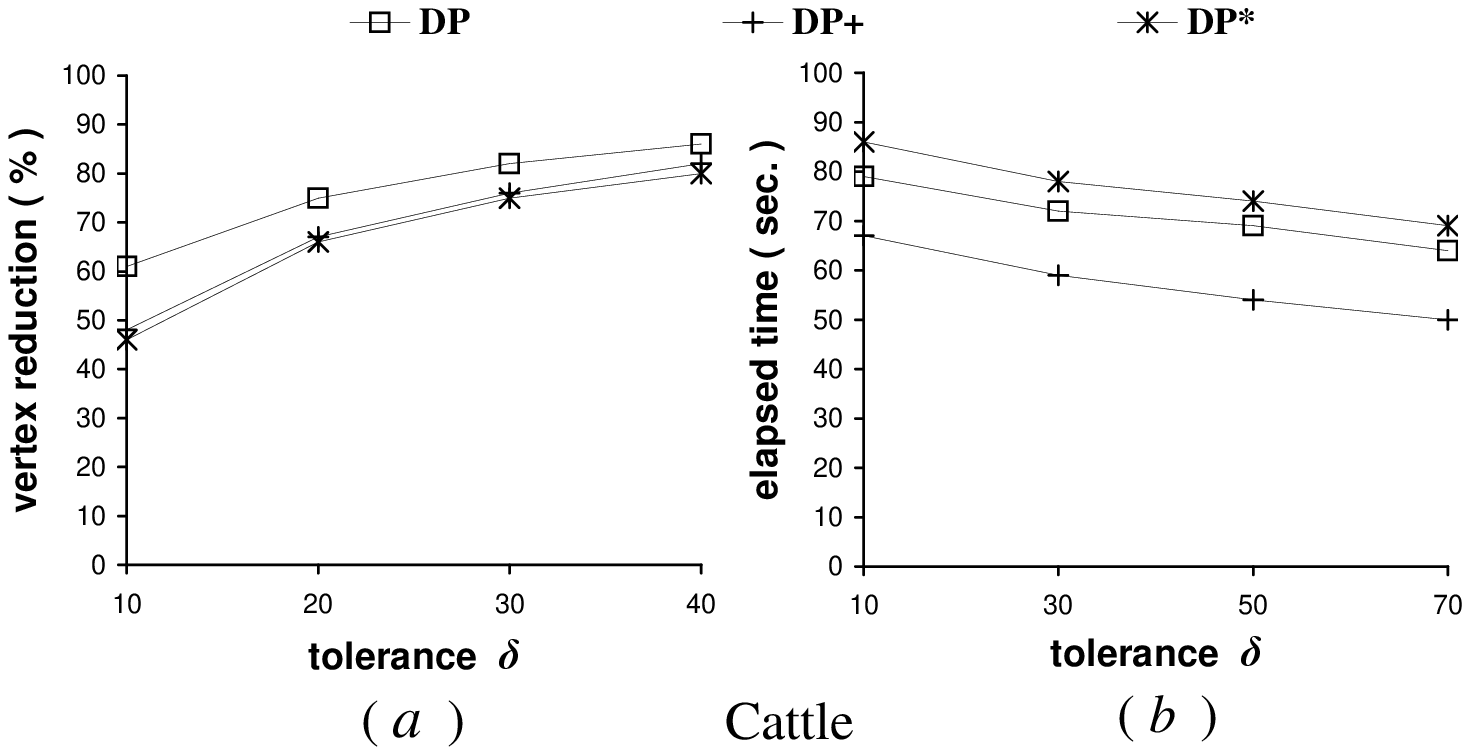}
    \caption{Comparison of Trajectory Simplification Methods}
    \label{exp_dp}
\end{figure}

Figure \ref{exp_dp}(\emph{a}) illustrates the differences of their reduction power for the Cattle
dataset. We skip the results for the other datasets because they show similar trends. With the same
values of tolerance, DP shows higher reduction rates than does DP*. This is natural since DP* uses
the time-ratio distance to approximate points, which is always equal to or greater than the
perpendicular distance of DP (see Section~\ref{sec:cuts*}). Furthermore, the vertex reduction of
DP+ is lower than that of DP. This is because DP+ does not preserve the shapes of the original
trajectories well when compared to DP. This aspect was explained in Section~\ref{sec:cuts+}.

\vspace{0.2cm}

In Figure \ref{exp_dp}(\emph{b}), DP+ exhibits the fastest elapsed time among the methods because
of its more effective division process. An interesting observation of the figure is that the
efficiency of all the methods grows as reduction ratios increase. Recall that all the methods
utilize the divide-and-conquer paradigm, which divides an input trajectory until no point exceeds a
given $\delta$. With a larger value of $\delta$, their division processes are likely to meet the
`end' quicker. For this reason, DP* also performs slower than the other methods (lower reduction
power than the others).

Next, we compare the discovery effectiveness and efficiency for the CuTS family. Given very large
values of $e$ and $\delta$, the CuTS family may produce one candidate containing all actual results
after the filter step and then the candidate may be divided into a large number of real convoys
through the refinement step. Thus, we cannot use the count of false positives as a measure of the
filters' effectiveness for our study.

Instead, we calculate \emph{refinement unit} that represents the computational cost of candidates
for the refinement step, which reflects the filtering power of each method effectively.
Specifically, the clustering cost of the convoy objects in each candidate is computed and then
multiplied by the candidate's lifetime. As mentioned earlier, the computational cost of clustering
is either $O(N^2)$ without index or $O(N \cdot \log N)$ with a spatial index. To clarify the
differences of each filter method, we considered the clustering without index support in our
experiments. For example, if a convoy candidate has 3 objects and its lifetime is 2, the refinement
unit is $3^2 \times 2 = 18$. Next, we aggregate each candidate's unit to obtain the total
refinement unit.

\begin{figure}[hbt]
    \center
    \includegraphics[width=8.5cm]{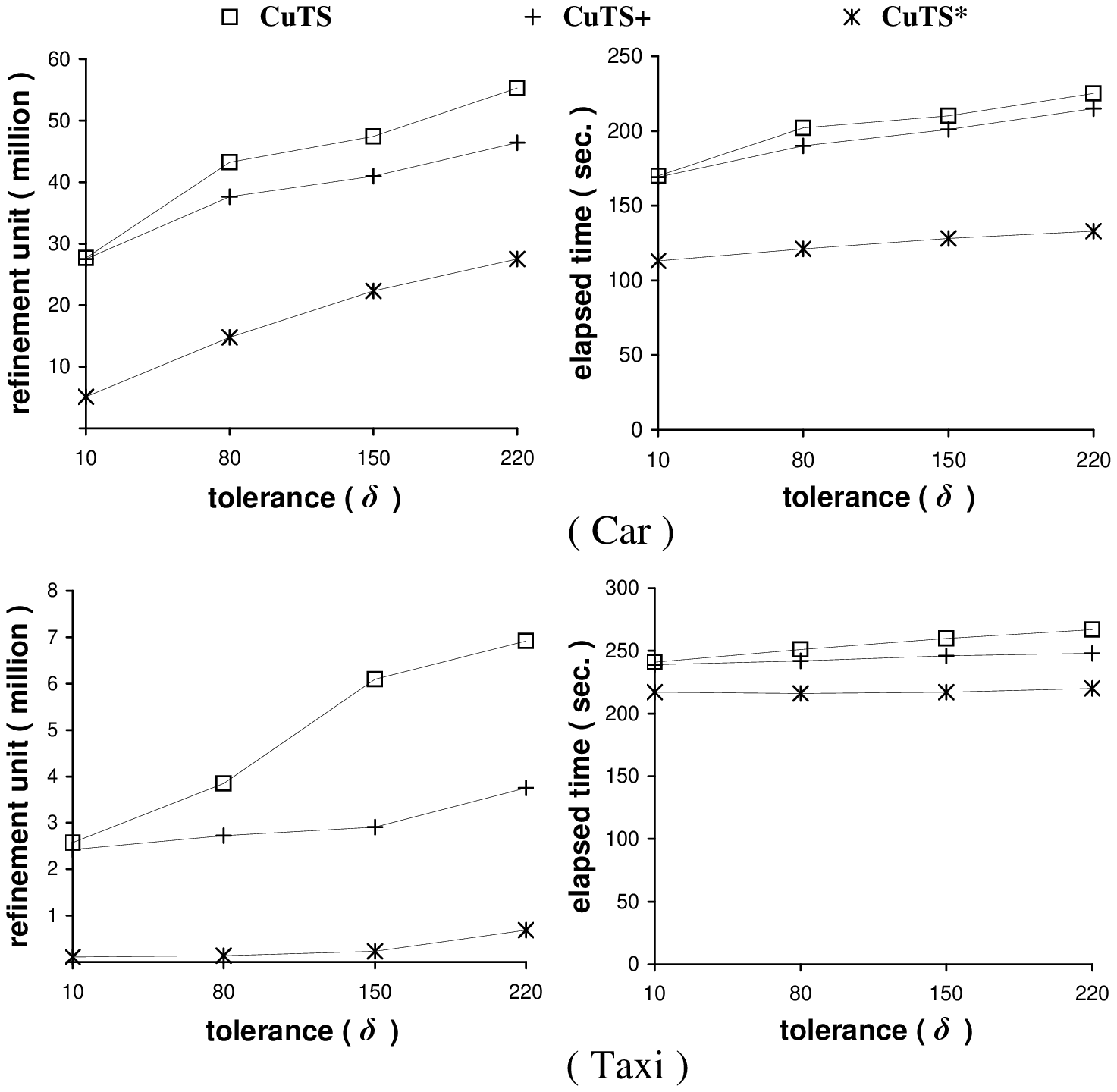}
    \caption{Effect of Simplification Tolerance ($\delta$)}
    \label{exp_tol}
\end{figure}

Figure~\ref{exp_tol} demonstrates the filtering power and the total discovery times for the CuTS
family when varying $\delta$. We omit the results for the Truck and Cattle datasets, but those two
datasets will be used in the next experiments. As expected, CuTS* has the lowest refinement unit
for both datasets, which yields the highest efficiency as well. In addition, CuTS+ has a better
filtering effectiveness than does CuTS. As discussed in Section~\ref{sec:cuts+}, the actual
tolerances obtained by DP+ of CuTS+ are always smaller or equal to those obtained by DP of CuTS. As
a result, the search range for clustering is reduced, and the filtering power grows in the figure.

\pagebreak

Another observation found in Figure~\ref{exp_tol}, for all members of the CuTS family is that both
the filters' effectiveness and the discovery efficiency decrease as the tolerance value increases
as the $\delta$ values affect not only the result of trajectory simplification, but also that of
range search for clustering.

Although the total elapsed times of the Car data grow steadily with increasing $\delta$, those of
the Taxi data stay almost constant or increase only very slightly. This is because the enlargement
of the search range is not sufficient to find more actual convoys with respect to the given
parameters. From this point, we can infer that the trajectories of the Taxi dataset are distributed
relatively uniformly, and thus the number of taxis traveling together within a given (reasonable)
distance is low.

\begin{figure}[hbt]
    \center
    \includegraphics[width=8.5cm]{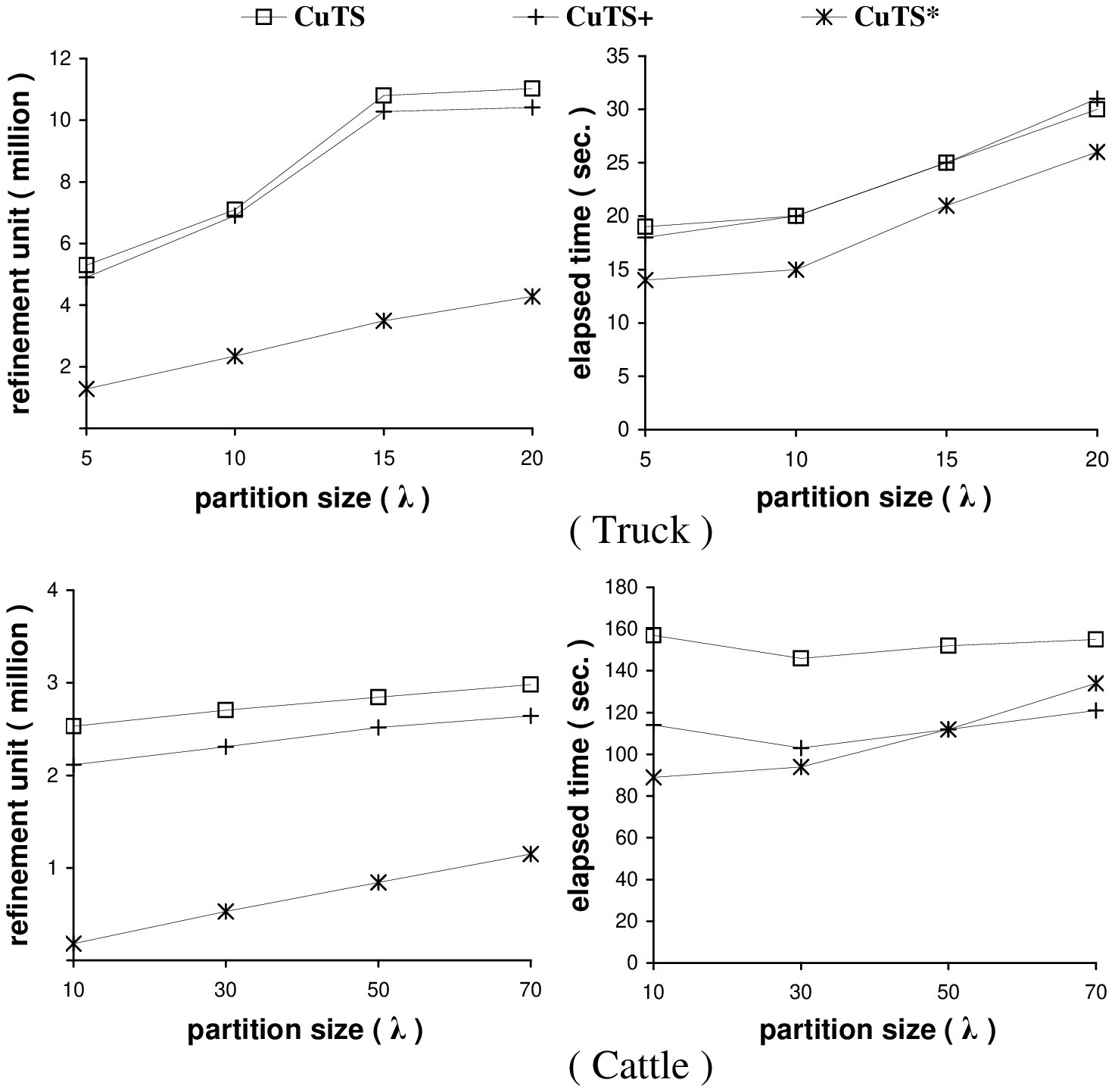}
    \caption{Effect of Time Partitioning ($\lambda$)}
    \label{exp_lambda}
\end{figure}

Lastly, we study how the size of the time partition $\lambda$ affects the results of the convoy
discovery. In fact, a large value of $\lambda$ yields an ineffective filtering step, whereas more
times of clustering are performed with a small value of $\lambda$ (see Section~\ref{sec:adaptive}).
In the Truck dataset of Figure~\ref{exp_lambda}, CuTS* shows better performance than the other
methods regardless of the $\lambda$ value. Also, both the effectiveness of the filters and the
efficiency of the discovery process decrease when $\lambda> 10$ for this dataset, for all methods
of the CuTS family.

On the other hand, the discovery efficiency of the CuTS family declines over the Cattle dataset
when $\lambda < 30$, although their refinement unit increases steadily in the same range of
$\lambda$. This implies that an appropriate $\lambda$ value is influenced by not only the filter's
effectiveness, but also another fact, possibly the length of trajectories since the average size of
Cattle's trajectories is very large.

Another interesting observation found in the Cattle dataset is that CuTS+ has similar efficiency to
CuTS*, and it is even faster for $\lambda \geq 50$. As seen in Figure~\ref{exp_anal}, trajectory
simplification is the key part of the total discovery time on this dataset. Therefore, faster
trajectory simplification (i.e., DP+) plays a more important role in the discovery efficiency in
this case.

\subsection{Parameter Determination of CuTS} \label{sec:param}

Proper values of $\delta$ and $\lambda$ may be difficult to find in some applications since they
are dependent on the data characteristics. In this section, we provide guidelines for determining
settings for these parameters. Note that the parameters do not affect the correctness of discovery
results, but only affect execution times.

%
%


\vspace{0.2cm} \noindent \textbf{Tolerance for trajectory simplification ($\delta$)} :  It is
obvious that a larger value of $\delta$ for DP of CuTS achieves a higher reduction result of
trajectory simplification. On the other hand, a large $\delta$ value is also used for the range
search of clustering in the CuTS algorithm; hence, the filter step of CuTS may not be tight enough
to prune many unnecessary candidate objects. In this tradeoff, our goal is to find a value
satisfying the following conditions : (i) the original trajectories become well simplified, and
(ii) the distance bounds are sufficiently tight, implying an effective filter process.

As the first step, we perform the original DP algorithm over a trajectory with $\delta=0$. In each
step of the division process (see details in Sections~\ref{sec:trjsimple} and \ref{sec:simple}), we
store the actual tolerance values in ascending order. Since $\delta=0$, the process continues until
all intermediate points of the original trajectory are tested.

\begin{figure}[hbt]
    \center
    \includegraphics[width=8.5cm]{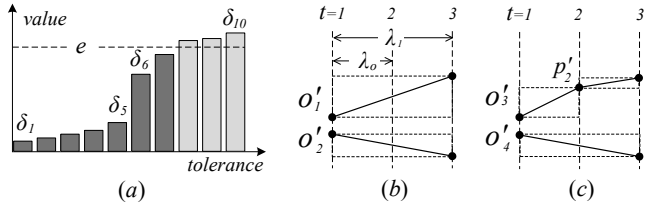}
    \caption{Value Selection of $\delta$ and $\lambda$}
    \label{fig_param}
\end{figure}

In the next step, we find the largest variance between two adjacent tolerances stored, and then
select the smaller one of those two tolerances. For example, assume that the DP method with
$\delta=0$ results in the 10 actual tolerance values $\delta_1,\delta_2,\cdots,\delta_{10}$ in
Figure~\ref{fig_param}(\emph{a}) through the first step. The difference in the tolerance values is
the largest between $\delta_5$ and $\delta_6$. We then select $\delta_5$ as a tolerance value
$\delta_s$. This selection is performed as long as $\delta_i < e$ (the dark gray bars in
Figure~\ref{fig_param}(\emph{a})). From our experimental studies, we found out that the filtering
power of the CuTS family decreases considerably on some datasets when we pick $\delta_i> e$.

Lastly, we perform the above steps for a sufficient time (e.g., 10\% of $N$) and average the
$\delta_s$ values selected to obtain a final $\delta$ for the processing of trajectory
simplification.

The idea behind this method is to find a relatively small $\delta$ value that achieves a reasonable
reduction through simplification. In the figure, if we pick $\delta_{10}$ and apply it to the
trajectory simplification, the reduction ratio will be nearly 100\%. Likewise, the use of
$\delta_5$ for the simplification is able to yield around a 50\% reduction although it does not
necessarily follow the same division processes with $\delta=0$ as the first step. If we pick
$\delta_6$ instead, it may bring (approximately) 60\% of trajectory reduction, which is slightly
higher than 50\%. However, the value of $\delta_6$ is much bigger than $\delta_5$, and the
effectiveness of range search can decrease dramatically.
%
%

\noindent \textbf{Length of time partition ($\lambda$)} : In Section~\ref{sec:adaptive}, we
discussed about dividing the time domain $T$ into time partitions for discovery processing, each of
which has length $\lambda$. If a time partition $\mathcal{T}_i$ has a large value for $\lambda$,
many line segments of a simplified trajectory within $\mathcal{T}_i$ form a long polyline. Thus,
the distances among those polylines become small, and many objects are likely to form a cluster
together, leading to ineffective filtering results. In contrast, a small value of $\lambda$
involves many computationally expensive clustering processes ($T/\lambda$ times).

Suppose that $o'_1$ and $o'_2$ in Figure~\ref{fig_param}(\emph{b}) are simplified trajectories. In
the figure, one clustering with $\lambda_1$ is obviously more efficient than two processes with
$\lambda_0$ because both cases have the same minimum distance between $o'_1$ and $o'_2$. From this
example, we can infer the value of $\lambda_1$ by computing $\frac{|o'|}{|o|} \times o.\tau$, where
$|o|$ ($|o'|$) is the number of points in the trajectory $o$ ($o'$) and $o.\tau$ is the time
interval of $o$.

In practice, however, there may be some time points that one (simplified) trajectory has, but
others do not have, such as $p'_2$ on $o'_3$ in Figure~\ref{fig_param}(\emph{c}). Using the
$\lambda_1$ for this case should not keep the filter's `good' effectiveness, and we need to lower
the $\lambda_1$ value. We can roughly estimate the probability that such case occurs by looking at
how densely a trajectory exits in the time space $T$. Notice that each trajectory may have a
different length ($o.\tau$) and may appear and disappear at any arbitrary time points in $T$. Thus,
the density of the trajectory is obtained by $o.\tau / T$. Finally, the probability that an object
has an intermediate time point within $\lambda_1$ is $(\lambda_1 - 2) \times o.\tau / T$. Together,
we obtain $\lambda = \lambda_1 - (\lambda_1 - 2) \times o.\tau / T$, rewriting $\lambda = o.\tau
\times ( \frac{|o'|}{|o|} \times ( 1 - \frac{o.\tau}{T} ) + \frac{2}{T})$.

So far, we have considered the computation of $\lambda$ for a single object. To obtain an overall
value of $\lambda$, we perform the above computation for all objects and average the values. Note
that all the statistics for this $\lambda$ computation can be easily gathered when a dataset is
loaded into the system (or one scan for disk-based implementations).

Although this method does not capture the distribution of a dataset precisely, the value of
$\lambda$ is quickly obtained and brings reasonable efficiency of the CuTS family.

\section{Conclusion} \label{sec:conc}
Discovering convoys in trajectory data is a challenging problem, and existing solutions to related
problems are ineffective at finding convoys. This study formally defines a \emph{convoy query}
using density-based notions, and it proposes four algorithms for computing the convoy query. Our
main algorithms (CuTS, CuTS+, and CuTS*) use line simplification methods as the foundation for a
filtering step that effectively reduces the amounts of data that need further processing. In order
to ensure that the filters do not eliminate convoys, we bound the errors of the discovery
processing over the simplified trajectories. Through our experimental results with real datasets,
we found that CuTS* showes the best performance. CuTS+ also performes well when the given
trajectories have a small number of objects and long histories.

\small
\section*{Acknowledgment}
National ICT Australia is funded by the Australian Government's Backing Australia's Ability
initiative, in part through the Australian Research Council (ARC). This work is supported by grant
DP0663272 from ARC.

\bibliographystyle{abbrv}

\pagebreak
\appendix
\normalsize

\section{Proofs of Lemmas}

\subsection{Proof of Lemma~\ref{lmm:prune-range}}

Consider the example of Figure~\ref{fig_proof}. To prove the lemma by contradiction, assume the
following equation holds:
%
\[
D(o_q(t),o_i(t)) \le e \label{eqn:1}
\]

Since $l'_q$ is a line segment (with actual tolerance
  $\delta(l'_q)$) in the simplified trajectory $o'_q$, there exists a
  location $a_q$ on $l'_q$ such that $D(a_q,o_q(t)) \le \delta(l'_q)$.
  Similarly, there exists a location $a_i$ on $l'_i$ such that
  $D(a_i,o_i(t)) \le \delta(l'_i)$. Due to the triangular inequality,
%
\[
D(a_q,a_i) \le D(a_q,o_q(t)) + D(o_q(t),o_i(t)) + D(o_i(t),a_i) \label{eqn:2}
\]

Combining the inequalities, we obtain:
\begin{eqnarray}
D(a_q,a_i) \le \delta(l'_q) + e + \delta(l'_i) \label{eqn:3}
\end{eqnarray}

On the other hand, $a_q$ ($a_i$) is a location on line segment
  $l'_q$ ($l'_i$). Hence, equation (\ref{eqn:4}) holds
\begin{eqnarray}
D_{LL}(l'_q,l'_i) \le D(a_q,a_i) \label{eqn:4}
\end{eqnarray}

From the last two inequalities (\ref{eqn:3}) and (\ref{eqn:4}), we get:
\begin{eqnarray}
D_{LL}(l'_q,l'_i) \le e +  \delta(l'_q) + \delta(l'_i) \label{eqn:5}
\end{eqnarray}

Therefore, the resulting contradiction of (\ref{eqn:5}) proves Lemma~\ref{lmm:prune-range}.

\subsection{Proof of Lemma~\ref{lmm:prune-group-range}}

Note that for all $l'_i \in S$, we have $\delta_{max}(S) \ge \delta(l'_i)$ and \\*
$D_{min}(\mathcal{B}(l'_q),\mathcal{B}(S)) \le D_{LL}(l'_q,l'_i)$. If the following equation
satisfies:
%
\[
D_{min}(\mathcal{B}(l'_q),\mathcal{B}(S)) > e + \delta(l'_q) + \delta_{max}(S)
\]
%
then, the next equation must also hold:
%
\[
D_{LL}(l'_q,l'_i) > e + \delta(l'_q) + \delta(l'_i)  \label{eqn:8}
\]

The rest of this proof follows directly from Lemma~\ref{lmm:prune-range}.

\subsection{Proof of Lemma~\ref{lmm:prune-star-range}}

Since $l'_q$ is a line segment (with actual tolerance $\delta(l'_q)$) in the simplified trajectory
$o'_q$ , the location $l'_q(t)$ meets:
%
\[
  D(l'_q(t),o_q(t)) \le \delta(l'_q) \label{eqn:9}
  \]

Similarly, the location $l'_i(t)$ satisfies:
%
\[
D(l'_i(t),o_i(t)) \le \delta(l'_i) \label{eqn:10}
\]

In addition, we have:
%
\[
D_{*}(l'_q,l'_i) \le D(l'_q(t),l'_i(t)) \label{eqn:11}
\]

The logic of the remainder of the proof is the same as in the proof of
  Lemma~\ref{lmm:prune-range}.

\pagebreak

\section{Additional Experiments}

%
%
%

\subsection{MC vs. CMC}\label{app:mc2}

In this experiment, we intend to demonstrate empirically that methods for the discovery of moving
clusters cannot be used to compute convoys directly (see Section~\ref{sec:trjclust}). Specifically,
we study the discovery accuracies of convoys by a solution for moving cluster (MC2). MC2 reports
results of the convoy query if the portion of common objects in any two consecutive clusters $c_1$
and $c_2$ is not below a given threshold parameter $\theta$, i.e., $\frac{ |c_1 \cap c_2| }{ |c_1
\cup c_2| } \ge \theta$.

\begin{figure}[hbt]
    \center
    \includegraphics[width=8.5cm]{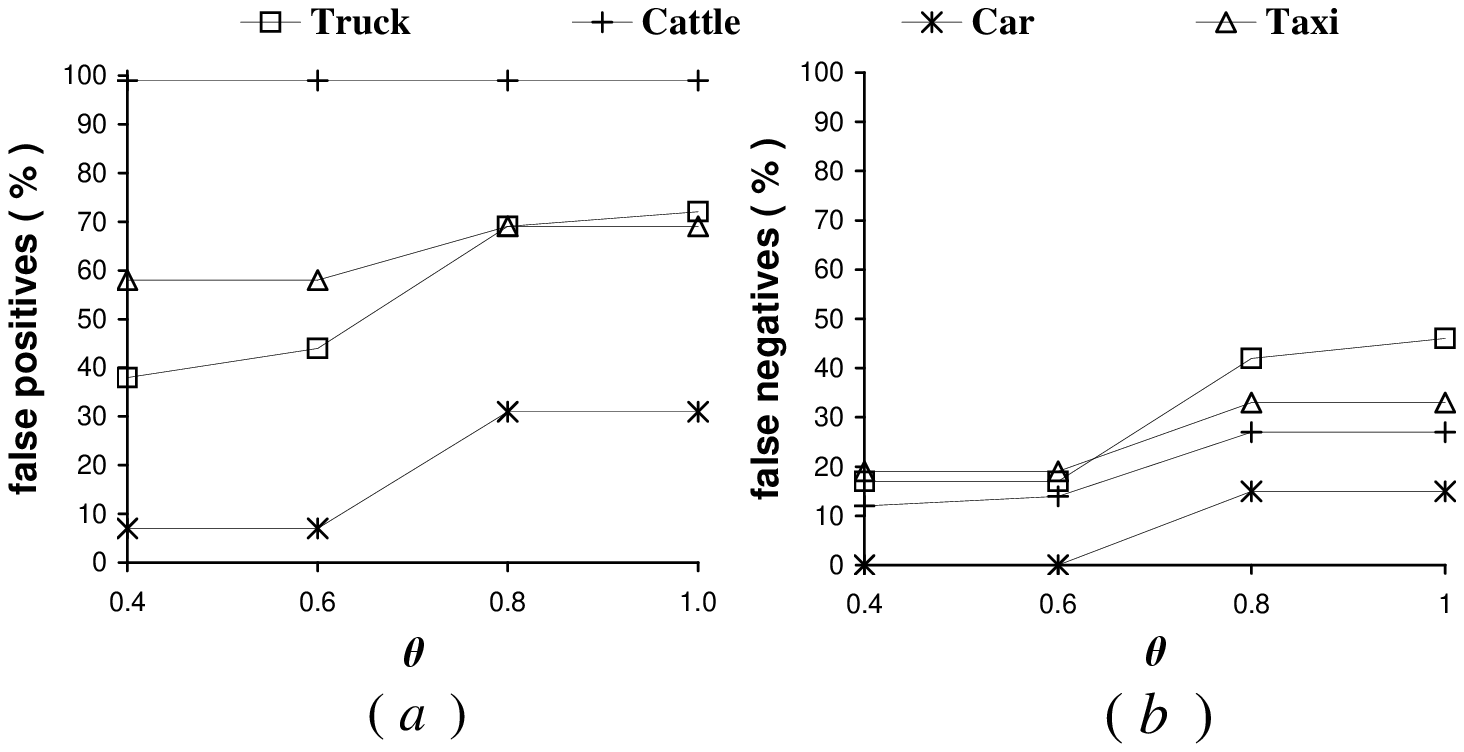}
    \caption{Discovery Quality of the MC method for Convoys}
    \label{exp_mc2}
\end{figure}

Let $R_m$ be a result set of convoys discovered by MC2 and $R_c$ be another set obtained by CMC (or
CuTS). We measure the proportions of false positives in Figure \ref{exp_mc2}(\emph{a}) by verifying
whether each convoy $v \in R_m$ satisfies the query condition with respect to $m$, $k$, and $e$
using the results of CMC (i.e., $(\frac{|R_m - R_c|}{|R_m|})\times 100$). Likewise, false negatives
in Figure \ref{exp_mc2}(\emph{b}) are computed by $(\frac{|R_c - R_m|}{|R_c|})\times 100$).

In fact, MC2 reported bigger numbers of convoys than what CMC does because MC2 does not have the
lifetime constraint $k$. This feature was especially obvious for the Cattle dataset that is larger
than the others. As a result, the proportions of actual convoys in the result set were very low,
and the numbers of false positives were very high in Figure \ref{exp_mc2}(\emph{a}). For the other
datasets, false positives went up as the $\theta$ value grew since the number of convoys reported
by MC2 also increased. Let $\theta_{c_1c_2}$ be a ratio of common objects between two snapshot
clusters $c_1$ and $c_2$. Assume that there are four consecutive snapshot clusters $c_1, c_2, c_3,$
and $c_4$, and $\theta_{c_1c_2}=1.0, \theta_{c_2c_3}=0.8, \theta_{c_3c_4}=1.0$. If we set the value
of $\theta$ to be equal to or smaller than 0.8, one moving cluster having all the snapshot clusters
will be reported (say $MC_{c_1c_2c_3c_4}$). In contrast, when $\theta > 0.8$, MC2 will discover two
moving clusters $MC_{c_1c_2}$ and $MC_{c_3c_4}$. Therefore, a higher $\theta$ value may produce a
larger number of moving clusters as convoy results.

Even though MC2 returns many convoys, the result set did not necessarily contain all actual
convoys. We investigate this aspect by computing false negatives in Figure \ref{exp_mc2}(\emph{b}).
In general, the number of false negatives increases as the $\theta$ value increases because the
number of convoys discovered by MC2 also increases.
%
Note that if many actual convoys exist for different parameter settings, the proportions of both
false positives and false negatives may increase considerably. Therefore, the use of moving cluster
methods for convoy discovery is ineffective and unreliable.

\end{document}